# Electrical Probing of Sub-Néel Spin Dynamics in Two-Dimensional Antiferromagnets Using Graphene Heterostructures

*Adrián García-Martín [a], Sara Gullace [a, *], Nicolás Montenegro-Pohlhammer [b], María José Martínez-Pérez [c], Rocío Sánchez-de-Armas [b], Carmen J. Calzado [b], Enrique Burzurí [a, d,*]*

[a] Dpto. Física de la Materia Condensada, Universidad Autónoma de Madrid, c/ Francisco Tomás y Valiente 7, 28049 Madrid, Spain.

[b] Dpto. Química Física, Universidad de Sevilla, c/Profesor García González, s/n, 41012 Sevilla, Spain.

[c] Instituto de Nanociencia y Materiales de Aragón (INMA), CSIC—Universidad de Zaragoza, Zaragoza, Spain.

[d] Condensed Matter Physics Center (IFIMAC) and Instituto Universitario de Ciencia de Materiales "Nicolás Cabrera" (INC), Universidad Autónoma de Madrid, c/ Francisco Tomás y Valiente 7, 28049 Madrid, Spain.

* sara.gullace@uam.es , enrique.burzuri@uam.es

**ABSTRACT**

Low-dimensional antiferromagnetic van der Waals materials have emerged as a versatile platform for exploring exotic spin dynamics and magnetic phases, yet electrically accessing these phenomena remains a major challenge because of their vanishing net magnetization and highly insulating nature. Here, we demonstrate that graphene can act as an ultrasensitive electrical transducer of hidden spin dynamics in two-dimensional antiferromagnets. By integrating graphene with the van der Waals antiferromagnet $FePS_3$, low-temperature transport measurements combined with magnetic-field and gate-voltage control reveal multiple resistance anomalies well below the Néel temperature. Their distinct temperature, magnetic-field, and carrier-density dependences enable us to associate these anomalies with magnon excitations and with a low-temperature magnetic reconfiguration of the antiferromagnetic state, both which remain largely inaccessible to conventional magnetometry. Density functional theory calculations further show that competing antiferromagnetic spin configurations in $FePS_3$ are nearly degenerate in energy while producing markedly different electronic responses in the adjacent graphene layer. The pronounced electrostatic tunability of these signatures demonstrates that graphene directly transduces interfacial magnetic dynamics into an electrical signal. Our work establishes graphene/antiferromagnetic van der Waals heterostructures as a versatile platform for the electrical readout of spin dynamics in two-dimensional magnets.

The confinement of magnetism to two dimensions (2D) profoundly modifies the balance between exchange interactions, magnetic anisotropy, and thermal fluctuations. Reduced dimensionality enhances both quantum and thermal fluctuations, promoting the emergence of magnetic phases and excitations that are absent or strongly suppressed in bulk materials.[1-4] Furthermore, symmetry breaking at surfaces and interfaces can enhance antisymmetric exchange interactions, such as the Dzyaloshinskii–Moriya interaction, thereby stabilizing topological spin textures and non-trivial magnetic states.[5] As a result, 2D magnetic systems have become a fertile platform for the exploration of extremely interesting phenomena;[2, 6] including skyrmions,[7] quantum spin liquids,[8] topological magnetic excitations,[9] and the quantum anomalous Hall effect.[10] Dimensional confinement also has a profound impact on spin dynamics. In two dimensions, magnons—the collective excitations of the spin lattice—play a central role in determining the stability of magnetic order and can themselves acquire non-trivial topological character.[11] The interplay between reduced dimensionality, magnetic anisotropy, and spin fluctuations gives rise to a rich landscape of magnetic phenomena that remains only partially understood.[12-14]

From a technological perspective, 2D magnetism offers unprecedented opportunities for device engineering.[15] The reduced screening and large surface-to-volume ratio facilitate the coupling of magnetism to external electric fields, enabling electrical control over key magnetic parameters such as exchange interactions,[16] critical temperatures,[17] magnetic anisotropy,[18] and electron–magnon coupling.[19] These properties make 2D magnetic materials particularly attractive for future spintronic and magnonic technologies.[2]

The discovery of van der Waals (vdW) magnetic materials has provided a major boost to the field by enabling the isolation of atomically thin magnetic crystals.[1] Since then, a wide variety of 2D magnets spanning different magnetic anisotropies, exchange mechanisms, and ordering temperatures have been identified and readily integrated into electronic and spintronic devices.[4, 20] Moreover, the vdW nature of these materials allows the fabrication of heterostructures in which interlayer coupling can be engineered with atomic precision.[21] This capability has opened new directions including moiré magnetism,[22] twisted magnetic heterostructures,[23] and magnetic twistronics.[24]

To date, however, most experimental advances have been achieved in ferromagnetic vdW materials, while antiferromagnetic systems have received comparatively less attention. This imbalance originates in part from the difficulty of detecting antiferromagnetic order in atomically thin crystals.[25] Many of the techniques that proved instrumental for the development of 2D magnetism, particularly magneto-optical probes, rely on the presence of a finite net magnetization and therefore become ineffective in compensated antiferromagnets. In addition, the development of conventional spintronics has historically been driven by ferromagnetic phenomena such as magnetoresistance and spin-valve effects, naturally biasing the emerging 2D field toward ferromagnetic materials.[26]

This situation has been changing rapidly in recent years. Antiferromagnetic vdW materials are increasingly recognized as a promising platform owing to the absence of stray fields, ultrafast spin dynamics, enhanced quantum fluctuations, and a broader diversity of magnetic ground states.[27, 28-31] In contrast to ferromagnets, which are largely characterized by parallel spin alignment, antiferromagnetic materials can host Néel, zigzag, stripy, armchair, and frustrated magnetic configurations, providing access to a significantly richer magnetic phase space.[20, 32] Despite this progress, probing the spin dynamics that develop well below the Néel temperature remains challenging. Most reports rely on spectroscopic techniques such as Raman scattering or neutron-based methods, while conventional superconducting quantum interference device (SQUID) magnetometry often lacks sensitivity to changes occurring within a fully compensated antiferromagnetic phase.[24, 33-35] Furthermore, these approaches are not naturally suited for electrical manipulation or integration into scalable device architectures. Electronic detection would constitute an attractive alternative; however, most magnetic vdW materials are highly insulating, preventing direct transport measurements.

Here we demonstrate that graphene can act as an ultrasensitive electrical transducer of hidden sub-Néel spin dynamics in van der Waals antiferromagnets. To this end, we fabricate and characterize field-effect transistors consisting of monolayer graphene coupled to the antiferromagnetic iron phosphorus trisulfide $FePS_3$. This material provides an ideal platform for the present study, as we have previously shown that its Néel transition produces an electrical signature in the adjacent graphene layer.[36] By combining low-temperature transport measurements with magnetic-field and electrostatic gate control, we uncover several distinct resistance anomalies in graphene occurring well below the Néel temperature, revealing a rich landscape of spin dynamics in $FePS_3$. Their evolution with temperature, magnetic field, and carrier density enables us to associate these anomalies with magnon excitations and with a transition between two antiferromagnetic configurations that remain largely invisible to conventional magnetometry. Complementary density functional theory calculations show that competing antiferromagnetic configurations in $FePS_3$ are nearly degenerate in energy while producing markedly different electronic responses in the adjacent graphene layer. Finally, the pronounced gate dependence of the observed signatures demonstrates that graphene does not merely sense the magnetic state of $FePS_3$, but electrically transduces its interfacial spin dynamics into a measurable transport signal.

Our results establish graphene/antiferromagnet van der Waals heterostructures as a versatile platform for the electrical readout of hidden spin dynamics in two-dimensional magnets

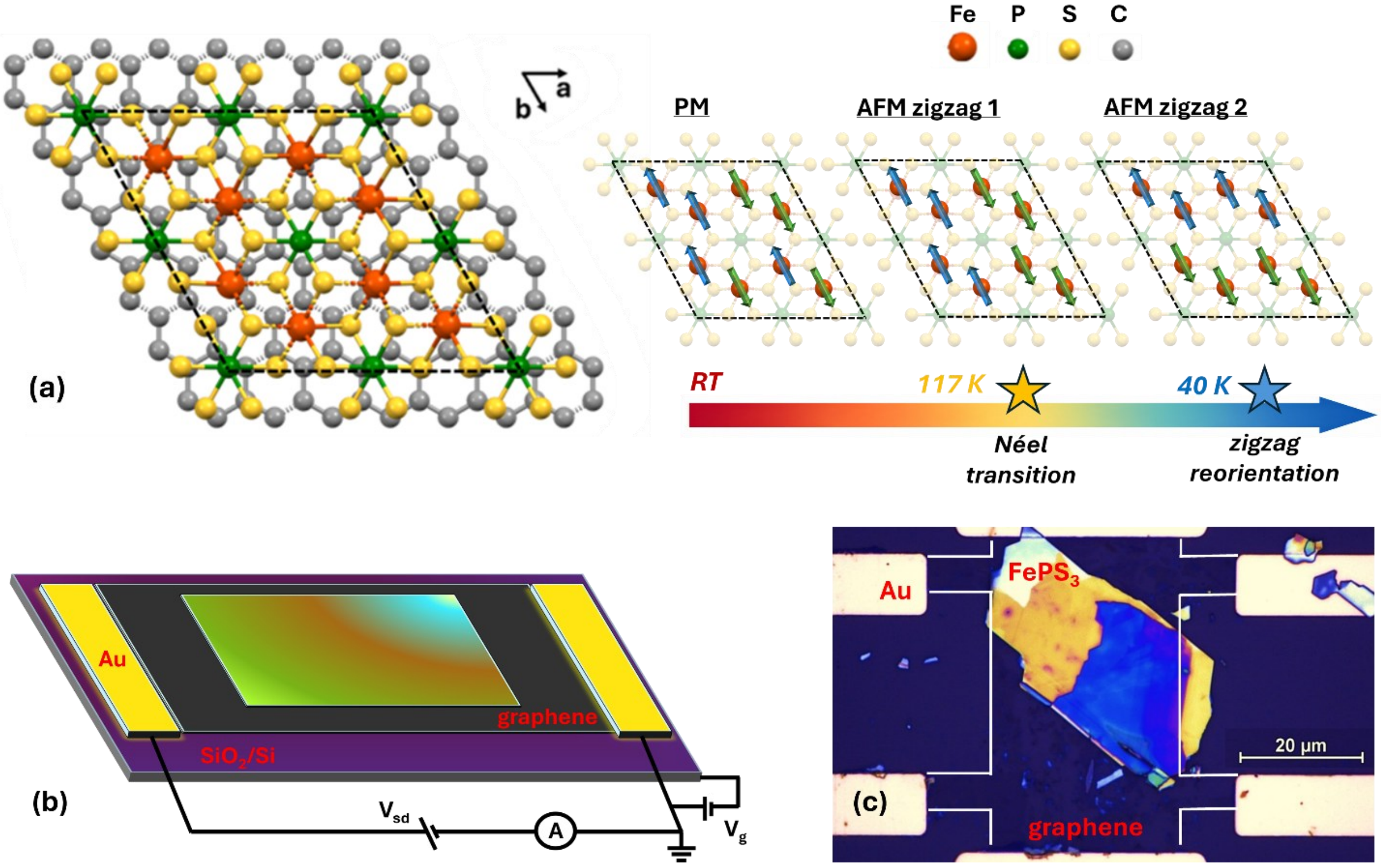

**Figure 1**. *(a)[Left] Atomistic top-view representation of a $FePS_3$ monolayer on top of a graphene monolayer as resulted from DFT calculations. [Right]Atomic level top-view representations of the magnetic structure of a $FePS_3$ monolayer, including the paramagnetic (PM) and two distinct antiferromagnetic (AFM) zigzag orders. Blue and green arrows on the magnetic Fe atoms represent the spin up and down, respectively. Yellow and blue stars mark the Néel transition and a reorientation of the zigzag AFM state with temperature. (b) Schematic representation of a $FePS_3$/Graphene Field-Effect Transistor (GFET) employed for the electron transport measurements. (c) Optical image of a representative $FePS_3$/GFET device. White lines highlight the underlying monolayer graphene in contact with the gold pads.*

The magnetic structure of iron phosphorus trisulfide $FePS_3$ consists of a 2D honeycomb lattice of Fe ions (S = 2) coupled via superexchange interactions. See atomistic top-view of $FePS_3$ in Figure 1a (atop a graphene monolayer as resulted from DFT calculations, see below). The magnetic ground state is predicted to be a zigzag out-of-plane Ising antiferromagnet where different zigzag alignments could, in principle, be accommodated, as seen in Figure 1a.[37] The paramagnetic-to-antiferromagnetic Néel transition occurs at $T_N$ = 118.3 K in bulk and slightly decreases for thinner flakes.[38] Figure 1b shows a representation of a $FePS_3$/chemical vapor deposited (CVD) graphene field-effect transistor heterostructure, hereafter FEPS/GFET, with a schematic view of the electrical circuit employed for the electron transport measurements. The devices are prepared by a dry transfer technique. Exfoliated $FePS_3$ flakes are initially obtained by repeated peeling of the bulk material with Nitto tape. Thereafter, the flakes are further exfoliated onto a piece of viscoelastic polydimethylsiloxane (PDMS) silicone polymer. The PDMS seal is mounted in a transfer station equipped with an optical microscope and a set of micromanipulators. The PDMS stamp area containing the selected flake is gently placed onto the CVDG window of a commercial CVDG FET. By carefully peeling off the PDMS seal, the $FePS_3$ stays adhered to the CVDG forming the FEPS/GFET heterostructure. See Fig. S1 in the Supporting Information for a summary of the fabrication process. The underlying $SiO_2$ (90 nm) and Si substrate (675 μm) are used as gate dielectric and gate electrode, respectively. Figure 1c shows an optical image of a representative FEPS/GFET. The magneto-electronic properties of the heterostructure are probed by measuring the drain-source electrical resistance ($R$) under different magnetic fields ($H$), temperatures ($T$) and gate voltages ($V_g$). The magnetic field is applied perpendicular to the device and flakes and therefore parallel to the $FePS_3$ Ising anisotropy axis. Note that metallic contacts are placed directly on graphene and that the $FePS_3$ resistance is orders of magnitude larger, therefore the current flows mainly through graphene.

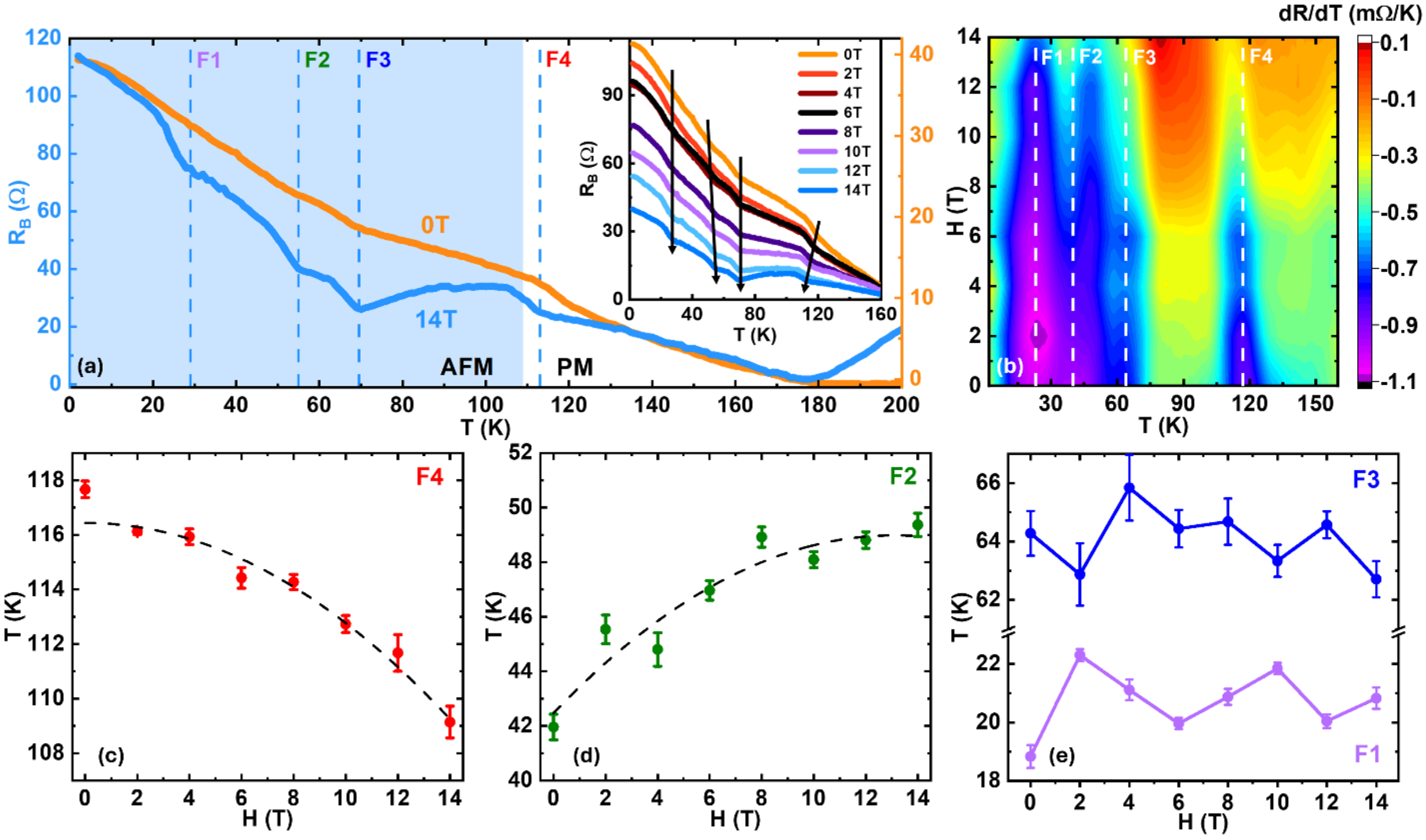

**Figure 2**. *(a) Resistance R measured on a FEPS/GFET heterostructure (device A) as a function of temperature T, at 0 T (orange) and 14 T (blue) applied magnetic field. Additional curves at intermediate magnetic fields are shown in the inset. Baselines based on minimum values are subtracted from the original resistance curves to facilitate comparison. Vertical dashed lines are visual help to localize anomalous features in the curve recorded at 14 T. These features are tagged as F1 (violet), F2 (green), F3 (dark blue) and F4 (red), respectively. (b) dR/dT color map obtained from (a) as a function of temperature and magnetic field. Individual dR/dT curves can be found in Figure S2 in the Supporting Information. The four main features (F1-4) are highlighted by white dashed lines at $T_F$ (H = 0). (c-e) Magnetic field-evolution of the characteristic temperature for (c)Feature 4, (d) Feature 2, and (e) Features 1 and 3. Vertical error bars are the error associated with the Lorentzian fittings in the original resistance derivative curves. The dashed black lines are fittings to phenomenological Equations 1 and 2 described in the main text.*

Figure 2a shows the resistance measured as a function of temperature for device A at zero external magnetic field (orange curve) and at $H = 14$ T (blue curve). Additional curves measured with an increasing applied magnetic field in the range 2-12 T are presented in the inset. When no external magnetic field is applied, the resistance of FEPS/GFET increases almost monotonically with decreasing temperature (orange curve) down to the lowest measured temperatures (~ 2K). A small bump appears at ∼117 K, indicative of the Néel transition to an antiferromagnetic state in $FePS_3$. The change in the magnetic state of the top layer is translated into a distinct charge transfer from $FePS_3$ into graphene, and thus changes in graphene resistance are detected, as reported before.[36] This bump strongly shifts to lower temperatures as the magnetic field increases (blue curve). Interestingly, a series of additional dips/peaks can be observed well below the Néel temperature at, approximately, $T$(F1) ~ 20 K, $T$(F2) ~ 40K, and $T$(F3) ~ 60K. These features, though faint at zero magnetic field, become more pronounced under applied magnetic fields (blue curve and curves in the inset of Figure 2a). None of these features are observed in bare graphene FETs without $FePS_3$, (device B, see Figures S3 and S4 in the Supporting Information), strongly suggesting that additional or secondary phenomena in $FePS_3$, appearing below the Néel temperature, are responsible for the emergence of these features, and that they can be electrically monitored through the graphene resistance. An additional sample (device C) displaying this behavior is shown in Figure S5 in the Supporting Information.

A color map of $dR/dT$ as a function of $T$ and $H$ (see Fig. 2b) facilitates the interpretation of the results. The four features can be clearly identified as dips in $dR/dT$ and fitted with Lorentzian curves at fixed applied magnetic field values (Figure S2 in the Supporting Information). This procedure allows us to extract the corresponding characteristic temperatures and investigate their evolution with magnetic field. Figures 2c-e show the field dependence of the four features. Interestingly, Feature 4, associated with the paramagnetic-to-antiferromagnetic transition, together with the newly observed Feature 2, exhibits a clear magnetic-field dependence. This behavior is expected for Feature 4, as the Néel temperature is known to decrease with increasing magnetic field. In this regime, the magnetic field competes with the antiferromagnetic exchange interactions, thereby destabilizing the antiferromagnetic phase and thus decreasing the Néel temperature. The magnetic field dependence of the Néel temperature $T_N$ is analyzed using a phenomenological scaling form:

$$T_N(H) = T_N(0) \cdot [1 - \left(\frac{H}{H_C}\right)^2] \quad (1)$$

where $H_C$ is an effective critical field associated with the suppression of antiferromagnetic order, so that $T_N(H_C) = 0$ K. The quadratic dependence on magnetic field corresponds to the leading-order correction allowed by symmetry, since the free energy must remain invariant under time reversal, forbidding linear terms in $H$. The experimental data are well described by this quadratic dependence with a characteristic field $H_C \sim 56$ T (dashed line in Figure 2c). The large value of $H_C$ indicates that the performed measurements remain far from the critical regime, justifying the use of the low expansion in Eq. 1. This result is in a relatively good agreement with pulsed high magnetic field measurements reported for $FePS_3$.[39] Such behavior is expected in systems with large magnetic anisotropy and magnetic fields aligned with the easy axis, as is the case for the $FePS_3$.

Feature 2, appearing at $T(H = 0 \text{ T}) \sim 42$ K, is particularly intriguing. A clear magnetic-field dependence is observed, with the corresponding dip in $dR/dT$ shifting to higher temperatures as the magnetic field increases, reaching $T(H = 14 \text{ T}) \sim 50$ K, as seen in Figure 2d. The positive field dependence strongly suggests that the anomaly occurring in $FePS_3$ at this temperature has a magnetic origin. However, no magnetic feature in this temperature range has been observed in conventional magnetometry measurements available in the literature.[38, 40] The magnetic field dependence cannot be fitted with a quadratic dependence but rather with a second order polynomial equation:

$$T(H) = a + b \cdot H + c \cdot H^2 \quad (2)$$

with fitting parameters, $a = 42.5 \pm 0.8$ K; $b = 0.99 \pm 0.24$ K/T and $c = -0.038$ K/T$^2$. Recent dielectric spectroscopy results on $FePS_3$ report an anomaly in its dielectric constant at approximately 40 K, [37] roughly the Feature 2 characteristic temperature observed in our electron transport measurements. This anomaly is explained in terms of a reorientation of the AFM-zigzag phase, involving a change in the preferential spin configuration from zigzag alignment along one crystallographic direction (zigzag1) to a perpendicular zigzag orientation (zigzag2), as seen in Figure 1a. This scenario is made possible by an in-plane structural anisotropy, recently reported and attributed to non-equivalent Fe-S bond lengths within the $FeS_6$ octahedron.[37] Although no definitive proofs have been provided for this scenario, it is consistent with some aspects of our observations. First, a reorientation between two antiferromagnetic states, with vanishing net magnetization, would produce little or no signature in SQUID measurements, whereas it may have a strong impact in the conductance of the FEPS/GFET device, as further supported by density functional theory (DFT) calculations below. Second, the observed magnetic-field dependence of the characteristic temperature suggests that the applied field stabilizes one zigzag configuration over the other. In this case, the response may not be strictly symmetric with respect to the field direction, allowing for lineal in $H$ terms in Eq. 2 that were not present in the paramagnetic-to-ferromagnetic transition (Feature 4). We thus tentatively ascribe Feature 2 to a phase reorientation between zigzag antiferromagnetic configurations sensed by graphene, where the magnetic field stabilizes the lowest energy configuration.

We now focus on the other two features in Figure 2e. Features 1 and 3 remain pinned at nearly the same characteristic temperature, regardless of the magnitude of the applied magnetic field. However, the resistance dip associated with both features becomes progressively more pronounced with increasing magnetic field (Figure 2a-b), pointing to a magnetic origin. Interestingly, recent Raman spectroscopy studies have identified a Raman-active mode associated with magnon excitations in $FePS_3$ appearing at 60K, [33, 34] which coincides with the characteristic temperature of Feature 3. In addition, magnon excitations appearing at 60 K were previously observed in the current-voltage characteristics of bare $FePS_3$ flakes.[38] Remarkably, in our heterostructures, these signatures appear in electrical transport mediated by the underlying graphene layer.

We therefore attribute Feature 3 to an additional transport contribution enabled by magnon excitations in $FePS_3$ and coupled to the electronic transport in graphene. The fact that the resistance minimum becomes progressively deeper under applied magnetic field indicates that this contribution is enhanced, rather than suppressed, by magnetic field. A possible microscopic origin could be an interfacial magnon–electron coupling mechanism, whereby thermally activated magnons in $FePS_3$ modify the scattering landscape or open an additional spin-mediated transport pathway in graphene. Within this picture, the resistance minimum, at approximately half the Néel temperature, emerges from the competition between the increasing population of magnetic excitations and the gradual loss of magnetic order at higher temperatures.

Finally, Feature 1 at the lowest temperature follows a similar trend to Feature 3, that is, no field dependence of the characteristic temperature (see Figure 2e) but a significant enhancement of the resistance step/dip with magnetic field. This is the only sub-Néel feature that has been observed in conventional SQUID measurements and is typically associated with spin dynamics like pinning of domain walls or freezing of magnonic modes, although no definitive proof has been provided.[38, 41]

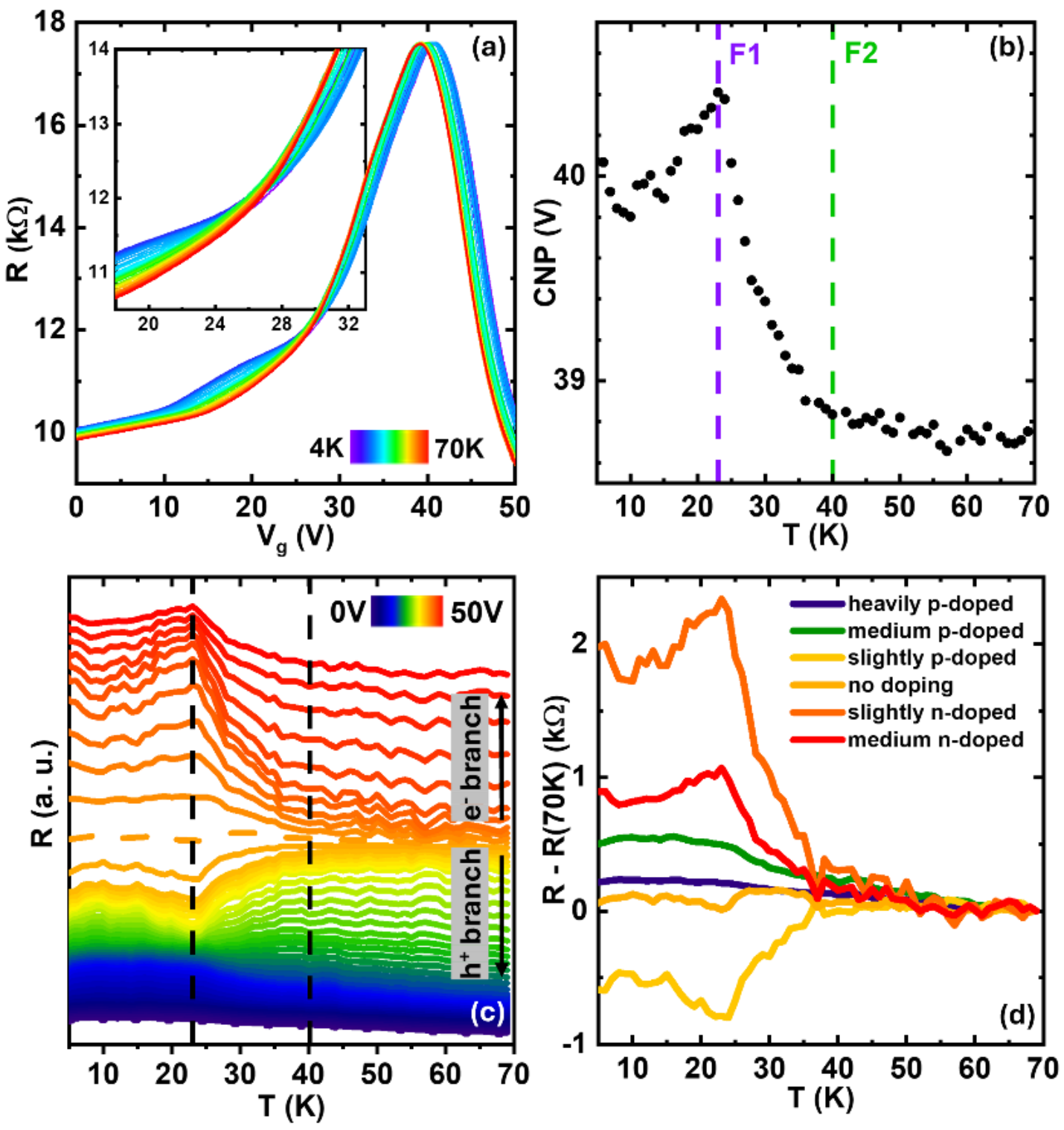

**Figure 3**. *(a) Gate voltage traces measured at temperatures from 4 K (blue) to 70 K (red), with a fixed bias voltage of 1 V (device D). Inset: zoom of a crossing area. (b) Charge neutrality point (CNP) as a function of temperature extracted from (a). Vertical dashed lines, coinciding with Feature 1 and Feature 2 are added as a visual guide indicating trend changes. (c) Resistance curves as a function of temperature at different gate voltage ranging from 0 V (blue) to 50 V (red). The curves are shifted vertically to facilitate the comparison. Vertical dashed lines mark Feature 1 and Feature 2. The curve corresponding to the charge neutrality gate voltage ∼40 V is plotted as an orange dashed line. That curve separates two different regions depending on the dominating charge carrier: electrons ($V_g$ > 40 V) and holes ($V_g$ < 40 V). (d) Selected curves from (c) corresponding to specific levels of doping. The curves are rescaled to the resistance at the highest temperature for a better comparison. The shape of the curves changes drastically at T ~ 40 K (Feature 2) depending on the specific charge carrier sign and density.*

To gain a deeper understanding of the observed features we have explored the graphene energy landscape by applying an external gate voltage ($V_g$), as the detection of these sub-Néel features may critically depend on the available charge states in graphene. The ohmic behavior of the devices (see Figure S6 in the Supporting Information) allows resistance values to be easily obtained. Figure 3a shows the resistance measured as function of $V_g$ at different temperatures below the Néel temperature. The bias voltage is fixed to 1V. The graphene characteristic Dirac cone-like, gate-dependent resistance curve can be observed. The charge neutrality point (CNP), that is, the maximum in resistance separating hole and electron branches, appears at $V_g \sim 38$ V in the FEPS/GFET at 70K. The initial positive $V_g$ value is usually ascribed to p-doping due to residues from the CVD growth, the device nanofabrication and initial charge transfer from $FePS_3$.[42, 43] Thereafter, two distinct features can be observed with temperature: (*i*) The CNP shifts to higher energies with decreasing temperature, and (*ii*) there is a non-trivial evolution of the hole branch with temperature, as seen in the inset of Figure 3a. Figure 3b shows the position of the CNP as a function of temperature, obtained from the measurements in Figure 3a. At high temperatures, the CNP position remains nearly constant at around $V_g = 38.8$ V. Interestingly, upon cooling below 40 K the CNP undergoes a sharp shift towards higher gate voltages, reaching $V_g = 40.4$ V. Upon further cooling, the CNP reverses its trend and shifts back towards lower gate voltages below approximately 23 K. These two characteristic temperatures coincide with Feature 2 and 1, respectively, observed in the magneto-electric study.

Further insight can be obtained by studying the temperature dependence of the resistance at different gate voltages, shown in Figure 3c. The orange dashed line marks the resistance curve corresponding to the CNP gate voltage ($V_g = 40.4$ V). Thus, all the curves above that line correspond to electron transport whereas all the curves below correspond to hole transport. Interestingly, the magnitude and even the sign of the resistance variation associated with Feature 2- the zigzag1-to-zigzag2 transition- dramatically depend on the carrier density. See Figure 3d for a few selected curves for clarity. For low and medium electron concentrations, the transition produces a pronounced increase in resistance, whereas for low hole concentrations it results in a marked decrease. At higher hole doping levels, the anomaly increases again in resistance, becomes progressively weaker, and eventually nearly undetectable. The latter case (dotted blue line in Figure 3c) well overlaps to $V_g$= 0 V measurements presented in Figure 2. These observations demonstrate that the coupling between the zigzag1-to-zigzag2 transition (Feature 2), as well as the low-temperature spin dynamics associated with Feature 1, and the charge carriers in graphene is highly sensitive to both carrier density and carrier type. Consequently, the ability of graphene to detect these magnetic phenomena can be strongly tuned electrostatically. These results are in sharp contrast with gate-dependent measurements in bare GFET (device E), that are basically unaffected by temperature, with the CNP remaining nearly constant at around 12.5 V in the entire temperature range (4-70K). Also, the temperature-dependent resistance curves do not present any anomaly as a function of carrier density and type. (See Figure S7 in the Supporting Information). Finally, Feature 3 is not clearly resolved in the zero-field measurements presented here.

We now explore the possible origin of Feature 2 by means of DFT calculations. The FEPS/GFET heterostructure has been modeled with a single layer of $FePS_3$ deposited on top of pristine graphene (See computational details in the Supporting Information).

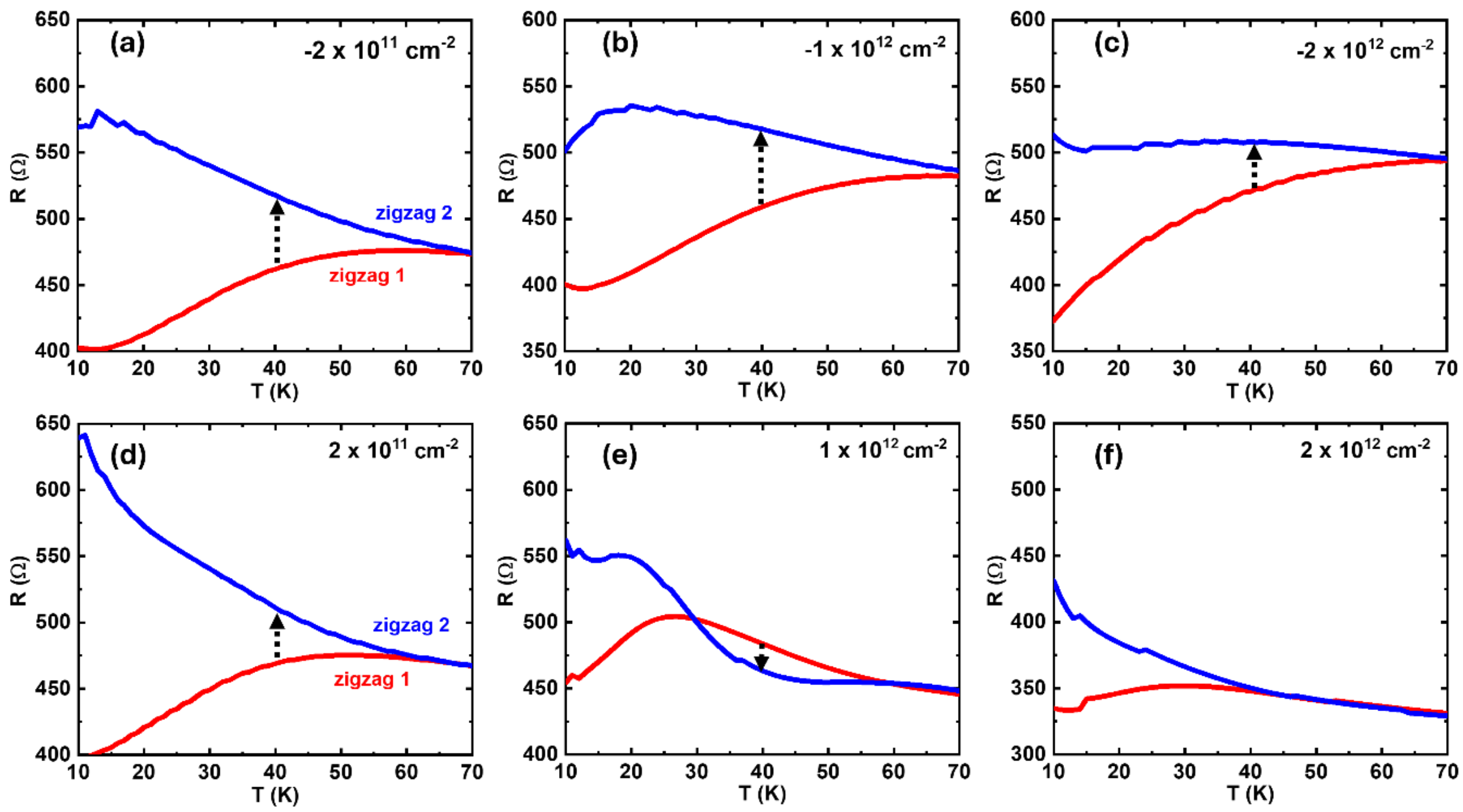

**Figure 4**. *Simulated transport properties for FEPS/GFET containing different initial (a)-(c) n-doping, and (d)-(f) p-doping concentrations. Red and blue curves represent two different zigzag AFM states. A black arrow is added as a visual guide for the secondary transitions happening between 23 and 40K. Complete curves are available in the Supporting Information.*

Two antiferromagnetic phases have been considered, zigzag 1 (antiferromagnetic zigzag in our previous study[36]) and zigzag 2 (Figure 1a), where the ferromagnetic chains are arranged along the *b* and *a* direction, respectively. These two zigzag phases are degenerate, both when considered as isolated layers and also in the presence of graphene. In fact, the zigzag1 phase is stabilized by just $2.5 \cdot 10^{-5}$ eV per Fe, in the case of the isolated layer, and $7.6 \cdot 10^{-5}$ eV per Fe center in the zigzag1/GFET heterostructure (Table S1). Interestingly, the unit cells of the optimized structure of the isolated $FePS_3$ antiferromagnetic zigzag layers are anisotropic in the *xy* plane, with unit cell parameters $a > b$ for zigzag1, and $a < b$ in the case of zigzag2. In both cases, the ferromagnetically aligned Fe chains run along the short bond direction. However, when deposited on graphene, the unit cell is isotropic in the *xy* plane, suggesting that the interaction with the substrate prevails over the magnetic order (Table S2). In other words, the magnetic interactions within $FePS_3$ are modulated by the presence of graphene, as also noted in our previous study dealing with the paramagnetic-to-antiferromagnetic transition at the Néel temperature.[36] Table S3 collects the Fe-Fe distances for both zigzag phases in isolated $FePS_3$ and in the FEPS/GFET heterostructures.

Strikingly, although energetically degenerate, the resulting FEPS/GFET heterostructures exhibit different transport properties at low temperatures (T < 70K). Figure 4 shows the simulated graphene resistivity in the heterostructure between 10 K and 70 K. The whole explored range of temperatures can be found in Figure S8 in the Supporting Information. Different p-type and n-type doping conditions have been considered, as it is known that graphene would be also intrinsically doped. Three different regimes can be distinguished around 40 K, where Feature 2 is observed. For n-type doping and slight p-doping ($\rho < 2\cdot10^{11}$ $cm^{-2}$, Figure 4a-d) the graphene resistivity is lower in the heterostructure with $FePS_3$ in the zigzag1 phase. Then the zigzag1-to-zigzag2 transition at 40 K produces an increment in resistivity upon cooling, as observed experimentally (Figure 3). For a medium p-type doping ($8\cdot10^{11}$ to $1.4\cdot10^{12}$ $cm^{-2}$, Figure 4e), the resistivity is higher for zigzag1. In this case, the zigzag1-to-zigzag2 transition results in a decrease of resistivity followed by its increase when cooling, in line with Figure 3c-d. Finally, for highly p-doped graphene (Figure 4f), both heterostructures behave similarly. Furthermore, no significant change is expected on resistivity when temperature decreases. For temperatures above 70 K the transport across these two heterostructures is almost indistinguishable.

To summarize, we fabricated and characterized field-effect transistors consisting of monolayer graphene coupled to the antiferromagnetic trichalcogenide $FePS_3$, aiming at the electrical probing of sub-Néel spin dynamics. By combining low-temperature transport measurements with magnetic-field and gate-voltage control, we uncovered several distinct resistance anomalies in graphene taking place well below the $FePS_3$ Néel temperature (~20, ~40 and ~60 K), revealing a complex landscape of spin dynamics in the antiferromagnet. The evolution of these features with temperature, magnetic field, and carrier density provides insight into their microscopic origin. In particular, we identified signatures associated with magnon excitations resulting in an additional spin-mediated transport pathway in graphene at approximately half the Néel temperature, and with a transition between two antiferromagnetic zigzag configurations, being well sensed by graphene, while remaining largely invisible with conventional magnetometry tools. Complementary DFT calculations showed that different antiferromagnetic states of $FePS_3$ can be nearly degenerate in energy while producing markedly different electronic responses in $FePS_3$/graphene heterostructures. Furthermore, gate-dependent measurements revealed a strong sensitivity of these phenomena to the carrier density in graphene, highlighting the interfacial nature of the coupling. The obtained results establish graphene as a highly sensitive electrical probe of antiferromagnetic spin dynamics and further demonstrate the potential of vdW heterostructures as ideal platforms for the investigation and control of complex magnetic phenomena in two dimensions.

## DATA AVAILABILITY STATEMENT

The data supporting the findings of this study are available within the paper and its Supporting Information. Additional data are available from the corresponding authors upon reasonable request.

## SUPPORTING INFORMATION

The Supporting Information is available free of charge at X.

It contains: Experimental methods, details about the DFT calculations and additional electron transport measurements on bare graphene and FEPS/GFET additional devices.

## ACKNOWLEDGMENTS

This work acknowledges financial support from the Spanish Ministry of Science and Innovation (MICINN) through the Ramón y Cajal fellowship RYC2019-028429-I (EB), the project CNS2024-154501(EB, SG), and the National Research Project PID2022-140923NB-C22 (EB, AG-M). Financial support through grant PID2024-161519NB-I00 funded by MICIU/AEI/10.13039/501100011033/ERDF/EU is acknowledged. The technical support of the Supercomputing Team of the Centro Informático Científico de Andalucía (CICA), and the access to the computational facilities of the "Centro de Servicios de Informática y Redes de Comunicaciones" (CSIRC, Universidad de Granada, Spain) and Cenits-COMPUTAEX (Extremadura, Spain) are deeply acknowledged.

## REFERENCES

1 Park, J.-G., Zhang, K., Cheong, H., Kim, J. H., Belvin, C., Hsieh, D., Ning, H., & Gedik, N. 2D van der Waals magnets: from fundamental physics to applications. *Rev. Mod. Phys.* **2025**, *98*, 025003. DOI: 10.48550/arXiv.2505.02355

2 Wang, Q. H., Bedoya-Pinto, A., Blei, M., Dismukes, A. H., Hamo, A., Jenkins, S., Koperski, M., Liu, Y., Sun, Q. C., Telford, E. J., Kim, H. H., Augustin, M., Vool, U., Yin, J. X., Li, L. H., Falin, A., Dean, C. R., Casanova, F., Evans, R. F. L., Chshiev, M., Mishchenko, A., Petrovic, C., He, R., Zhao, L., Tsen, A. W., Gerardot, B. D., Brotons-Gisbert, M., Guguchia, Z., Roy, X., Tongay, S., Wang, Z., Hasan, M. Z., Wrachtrup, J., Yacoby, A., Fert, A., Parkin, S., Novoselov, K. S., Dai, P., Balicas, L., & Santos, E. J.

G. The magnetic genome of two-dimensional van der Waals materials. *ACS Nano* **2022**, *16* (5), 6960–7079. DOI: 10.1021/acsnano.1c09150

3 Li, Y., Yang, B., Xu, S., Huang, B., Duan, W. Emergent phenomena in magnetic two-dimensional materials and van der Waals heterostructures. *ACS Applied Electronic Materials* **2022**, *4* (7), 3278-3302. DOI: 10.1021/acsaelm.2c00419

4 H. Li, S. Ruan & Y.-J. Zeng. Intrinsic van der Waals magnetic materials from bulk to the 2D limit: new frontiers of spintronics. *Adv. Mater*. **2019**, *31*(27), 1900065. DOI: 10.1002/adma.201900065

5 Fert, A., Cros, V., & Sampaio, J. Skyrmions on the track. *Nature Nanotechnology* **2013**, *8* (3), 152–156. DOI: 10.1038/nnano.2013.29

6 Grubišić-Čabo, A., Guimarães, M. H. D., Afanasiev, D., Garcia Aguilar, J. H., Aguilera, I., Ali, M. N., Bhattacharyya, S., Blanter, Y. M., Bosma, R., Cheng, Z., Dan, Z., Dash, S. P., Medina Dueñas, J., Fernandez-Rossier, J., Gibertini, M., Grytsiuk, S., Houmes, M. J. A., Isaeva, A., Knekna, C., Kole, A. H., Kurdi, S., Lado, J. L., Mañas-Valero, S., Lopes, J. M. J., Marian, D., Na, M., Pabst, F., Pierantoni, S. B., Regout, M., Reho, R., Rösner, M., Sanz, D., van der Sar, T., Sławińska, J., Verstraete, M. J., Waseem, M., van der Zant, H. S. J., Zanolli, Z. & Soriano, D. Roadmap on quantum magnetic materials. *2D Mater.* **2025**, *12*(3), 031501. DOI: 10.1088/2053-1583/adbe89

7 Li, X., Liu, X., Yang, J., Zhang, Y., & Pan, Y. Creation and manipulation of magnetic skyrmions in 2D van der Waals magnets. *Mat. Today Phys.* **2025**, *54*, 101727. DOI: 10.1016/j.mtphys.2025.101727

8 Broholm, C., Cava, R. J., Kivelson, S. A., Nocera, D. G., Norman, M. R., & Senthil, T. Quantum Spin Liquids. *Science* **2020**, *367* (6475). DOI: 10.1126/science.aay0668

9 Chen, L., Chung, J. H., Gao, B., Chen, T., Stone, M. B., Kolesnikov, A. I., Huang, Q., & Dai, P. Topological Spin Excitations in Honeycomb Ferromagnet $CrI_3$. *Phys. Rev. X* **2018**, *8* (4), 041028. DOI: 10.1103/PhysRevX.8.041028

10 Yujun Deng et al., Quantum anomalous Hall effect in intrinsic magnetic topological insulator $MnBi_2Te_4$. *Science* **2020**, *367* (6480), 895-900. DOI: 10.1126/science.aax8156

11 Lujan, D., Choe, J., Rodriguez-Vega, M., Ye, Z., Leonardo, A., Nunley, T. N., Chang, L. J., Lee, S. F., Yan, J., Fiete, G. A., He, R., & Li, X. Magnons and magnetic fluctuations in atomically thin $MnBi_2Te_4$. *Nat. Commun.* **2022**, *13* (2527). DOI: 10.1038/s41467-022-29996-w

12 Jiang, X., Liu, Q., Xing, J., Liu, N., Guo, Y., Liu, Z., & Zhao, J. Recent progress on 2D magnets: Fundamental mechanism, structural design and modification. *App. Phys. Rev.* **2021**, *8* (3), 031305. DOI: 10.1063/5.0039979

13 Kim, K., Lim, S. Y., Lee, J. U., Lee, S., Kim, T. Y., Park, K., Jeon, G. S., Park, C. H., Park, J. G., & Cheong, H. Suppression of magnetic ordering in XXZ-type antiferromagnetic monolayer $NiPS_3$. *Nat. Commun.* **2019**, *10* (345). DOI: 10.1038/s41467-018-08284-6

14 Y. Hou & R. Wu. Magnetic anisotropy in 2D van der Waals magnetic materials and their heterostructures: importance, mechanisms, and opportunities. *Adv. Funct. Mater.* **2025**, *35* (51), e09453. DOI: 10.1002/adfm.202509453

15 Marrows, C. H., Barker, J., Moore, T. A., & Moorsom, T. Neuromorphic computing with spintronics. *npj Spintronics* **2024**, *2* (12). DOI: 10.1038/s44306-024-00019-2

16 Mankovsky, S., Simon, E., Polesya, S., Marmodoro, A., & Ebert, H. Electric-field control of the exchange interactions. *Phys. Rev. B* **2021**, *104* (17), 174443. DOI: 10.1103/PhysRevB.104.174443

17 Shimamura, K., Chiba, D., Ono, S., Fukami, S., Ishiwata, N., Kawaguchi, M., Kobayashi, K., & Ono, T. Electrical control of Curie temperature in cobalt using an ionic liquid film. *App. Phys. Lett.*, **2012**, *100* (12), 122402. DOI: 10.1063/1.3695160

18 Yadav, A., Stojic, N. & Binggeli, N. Mechanism of electric-field tunable magnetic anisotropy in multiferroic $CuMP_2S_6$ (M=Cr, Mo) monolayers. *Phys. Rev. Materials* **2025**, *9* (12), 124409. DOI: 10.1103/yv1f-zrk2

19 Cheng, Y., Chen, K., & Zhang, S. Interplay of magnon and electron currents in magnetic heterostructure. *Physical Review B* **2017**, *96*(2), 024449. DOI: 10.1103/PhysRevB.96.024449

20 Rahman, S., Torres, J.F., Khan, A.R., & Lu, Y. Recent Developments in van der Waals Antiferromagnetic 2D Materials: Synthesis, Characterization, and Device Implementation. *ACS Nano* **2021**, *15* (11), 17175-17213. DOI: 10.1021/acsnano.1c06864

21 Bärenfänger, J., Zollner, K., Cvitkovich, L., Watanabe, K., Taniguchi, T., Hartl, S., Fabian, J., Eroms, J., Weiss, D., & Ciorga, M. Highly efficient lateral spin valve device

based on graphene/hBN/$Fe_3GeTe_2$. *2D Mater.* **2025**, *12*(4), 045008. DOI: 10.1088/2053-1583/adf453

22 Mellado, P. Magnetic moiré systems: a review. *Phys. Cond. Matt.* **2025**, *37*(32), 323001. DOI: 10.1088/1361-648X/adf483

23 Boix-Constant, C., Rybakov, A., Miranda-Pérez, C., Martínez-Carracedo, G., Ferrer, J., Mañas-Valero, S., & Coronado, E. Programmable Magnetic Hysteresis in Orthogonally-Twisted 2D CrSBr Magnets via Stacking Engineering. *Adv. Mater.* **2025**, *37*(8), 2415774. DOI: 10.1002/adma.202415774

24 Kim, K. M., Kiem, D. H., Bednik, G., Han, M. J., & Park, M. J. Ab Initio Spin Hamiltonian and Topological Noncentrosymmetric Magnetism in Twisted Bilayer $CrI_3$. *Nano Lett.* **2023**, *23*(13), 6088–6094. DOI: 10.1021/acs.nanolett.3c01529

25 Kumar, R., & Park, J. G. van der Waals antiferromagnets: From early discoveries to future directions in the 2D limit. *Journal of Magnetism and Magnetic Materials* **2026**, *645*, 173977. DOI: https://doi.org/10.1016/j.jmmm.2026.173977

26 Leitao, D. C., Riel, F. J. F. van, Rasly, M., Araujo, P. D. R., Salvador, M., Paz, E., & Koopmans, B. Enhanced performance and functionality in spintronic sensors. *npj Spintronics* **2024**, *2*(54). DOI: 10.1038/s44306-024-00058-9

27 Guo, Z., Wang, X., Wang, W., Zhang, G., Zhou, X., & Cheng, Z. Spin-Polarized Antiferromagnets for Spintronics. *Advanced Materials* **2025**, *37* (36), 2505779. DOI: 10.1002/adma.202505779

28 Dal Din, A., Amin, O. J., Wadley, P., & Edmonds, K. W. Antiferromagnetic spintronics and beyond. *npj Spintronics* **2024**, *2*(25). DOI: 10.1038/s44306-024-00029-0

29 Jia, Z., Zhao, M., Chen, Q., Tian, Y., Liu, L., Zhang, F., Zhang, D., Ji, Y., Camargo, B., Ye, K., Sun, R., Wang, Z., & Jiang, Y. Spintronic Devices upon 2D Magnetic Materials and Heterojunctions. *ACS Nano* **2025**, *19*(10), 9452–9483. DOI: 10.1021/acsnano.4c14168

30 Chen, J., Jin, Z., Yuan, R., Wang, H., Jia, H., Wei, W., Sheng, L., Wang, J., Zhang, Y., Liu, S., Yu, D., Ansermet, J. P., Yan, P., & Yu, H. Observation of Coherent Gapless Magnons in an Antiferromagnet. *Physical Review Letters* **2025**, *134*(5), 056701. DOI: 10.1103/PhysRevLett.134.056701

31 Yuan, H. Y., Yuan, Z., Duine, R. A., & Wang, X. R. Recent progress in antiferromagnetic dynamics. *EPL* **2020**, *132*(5), 57001. DOI: 10.1209/0295-5075/132/57001

32 Zhang, S., Xu, R., Luo, N., & Zou, X. Two-dimensional magnetic materials: Structures, properties and external controls. *Nanoscale* **2021**, *13* (3), 1398–1424. DOI: 10.1039/d0nr06813f

33 Ghosh, A., Palit, M., Maity, S., Dwij, V., Rana, S., & Datta, S. Spin-phonon coupling and magnon scattering in few-layer antiferromagnetic $FePS_3$. *Phys. Rev. B* **2021**, *103* (6), 064431. DOI: 10.1103/PhysRevB.103.064431

34 McCreary, A., Simpson, J. R., Mai, T. T., McMichael, R. D., Douglas, J. E., Butch, N., Dennis, C., Valdés Aguilar, R., & Hight Walker, A. R. Quasi-two-dimensional magnon identification in antiferromagnetic $FePS_3$ via magneto-Raman spectroscopy. *Phys. Rev. B* **2020**, *101* (6), 064416. DOI: 10.1103/PhysRevB.101.064416

35 Liao, J., Huang, Z., Shangguan, Y., Zhang, B., Cheng, S., Xu, H., Kajimoto, R., Kamazawa, K., Bao, S., & Wen, J. Spin and lattice dynamics in the van der Waals antiferromagnet $MnPSe_3$. *Phys. Rev. B* **2024**, *109* (22), 224411. DOI: 10.1103/PhysRevB.109.224411

36 García-Martín, A., Sánchez-De-Armas, R., Gullace, S., Calle, E., Martín-Pérez, L., Montenegro-Pohlhammer, N., Jaafar, M., Sanchez Costa, J., Calzado, C. J., & Burzurí, E. Conductance modulation in graphene/antiferromagnet van der Waals heterostructures induced by magnetic order. *Nanoscale* **2025**, *17* (35), 20327–20337. DOI: 10.1039/d5nr02538a

37 Ghosh, A., Birowska, M., Ghose, P. K., Rybak, M., Maity, S., Ghosh, S., Das, B., Dey, K., Bera, S., Bhardwaj, S., Nandi, S., & Datta, S. Anisotropic magnetodielectric coupling in layered antiferromagnetic $FePS_3$. *Phys. Rev. B* **2023**, *108*(6), L060403. DOI: 10.1103/PhysRevB.108.L060403

38 Martín-Pérez, L., Medina Rivero, S., Vázquez Sulleiro, M., Naranjo, A., Gómez, I. J., Ruíz-González, M. L., Castellanos-Gomez, A., Garcia-Hernandez, M., Pérez, E. M., & Burzurí, E. Direct magnetic evidence, functionalization, and low-temperature magneto-electron transport in liquid-phase exfoliated $FePS_3$. *ACS Nano* **2023**, *17* (3), 3007–3018. DOI: 10.1021/acsnano.2c11654

39 Wildes, A. R., Lançon, D., Chan, M. K., Weickert, F., Harrison, N., Simonet, V., Zhitomirsky, M. E., Gvozdikova, M. v., Ziman, T., & Rønnow, H. M. High field

magnetization of $FePS_3$. *Phys. Rev. B* **2020**, *101*(2), 024415. DOI: 10.1103/PhysRevB.101.024415

40 Lee, J. U., Lee, S., Ryoo, J. H., Kang, S., Kim, T. Y., Kim, P., Park, C. H., Park, J. G., & Cheong, H. Ising-type magnetic ordering in atomically thin $FePS_3$. *Nano Lett.* **2016**, *16* (12), 7433–7438. DOI: 10.1021/acs.nanolett.6b03052

41 Mayorga-Martinez, C. C., Sofer, Z., Sedmidubský, D., Huber, Š., Eng, A. Y. S., & Pumera, M. Layered metal thiophosphite materials: magnetic, electrochemical, and electronic properties. *ACS App. Mater. and Inter.* **2017**, *9* (14), 12563–12573. DOI: 10.1021/acsami.6b16553

42 Goniszewski, S., Adabi, M., Shaforost, O., Hanham, S. M., Hao, L., & Klein, N. Correlation of p-doping in CVD Graphene with Substrate Surface Charges. *Sci. Rep.* **2016**, *6*, 22858. DOI: 10.1038/srep22858

43 Pirkle, A., Chan, J., Venugopal, A., Hinojos, D., Magnuson, C. W., McDonnell, S., Colombo, L., Vogel, E. M., Ruoff, R. S., & Wallace, R. M. The effect of chemical residues on the physical and electrical properties of chemical vapor deposited graphene transferred to $SiO_2$. *App. Phys. Lett.* **2011**, *99* (12), 122108. DOI: 10.1063/1.3643444

# Supporting Information

for

# Electrical Probing of Sub-Néel Spin Dynamics in Two-Dimensional Antiferromagnets Using Graphene Heterostructures

*Adrián García-Martín [a], Sara Gullace [a*], Nicolás Montenegro-Pohlhammer [b], María José Martínez-Pérez[c], Rocío Sánchez-de-Armas [b], Carmen J. Calzado [b], Enrique Burzurí [a, d,*]*

[a] Dpto. Física de la Materia Condensada, Universidad Autónoma de Madrid, c/ Francisco Tomás y Valiente 7, 28049 Madrid, Spain.

[b] Dpto. Química Física, Universidad de Sevilla, c/Profesor García González, s/n, 41012 Sevilla, Spain

[c]Instituto de Nanociencia y Materiales de Aragón (INMA), CSIC—Universidad de Zaragoza, Zaragoza, Spain.

[d] Condensed Matter Physics Center (IFIMAC) and Instituto Universitario de Ciencia de Materiales “Nicolás Cabrera” (INC), Universidad Autónoma de Madrid, c/ Francisco Tomás y Valiente 7, 28049 Madrid, Spain.

* sara.gullace@uam.es  enrique.burzuri@uam.es

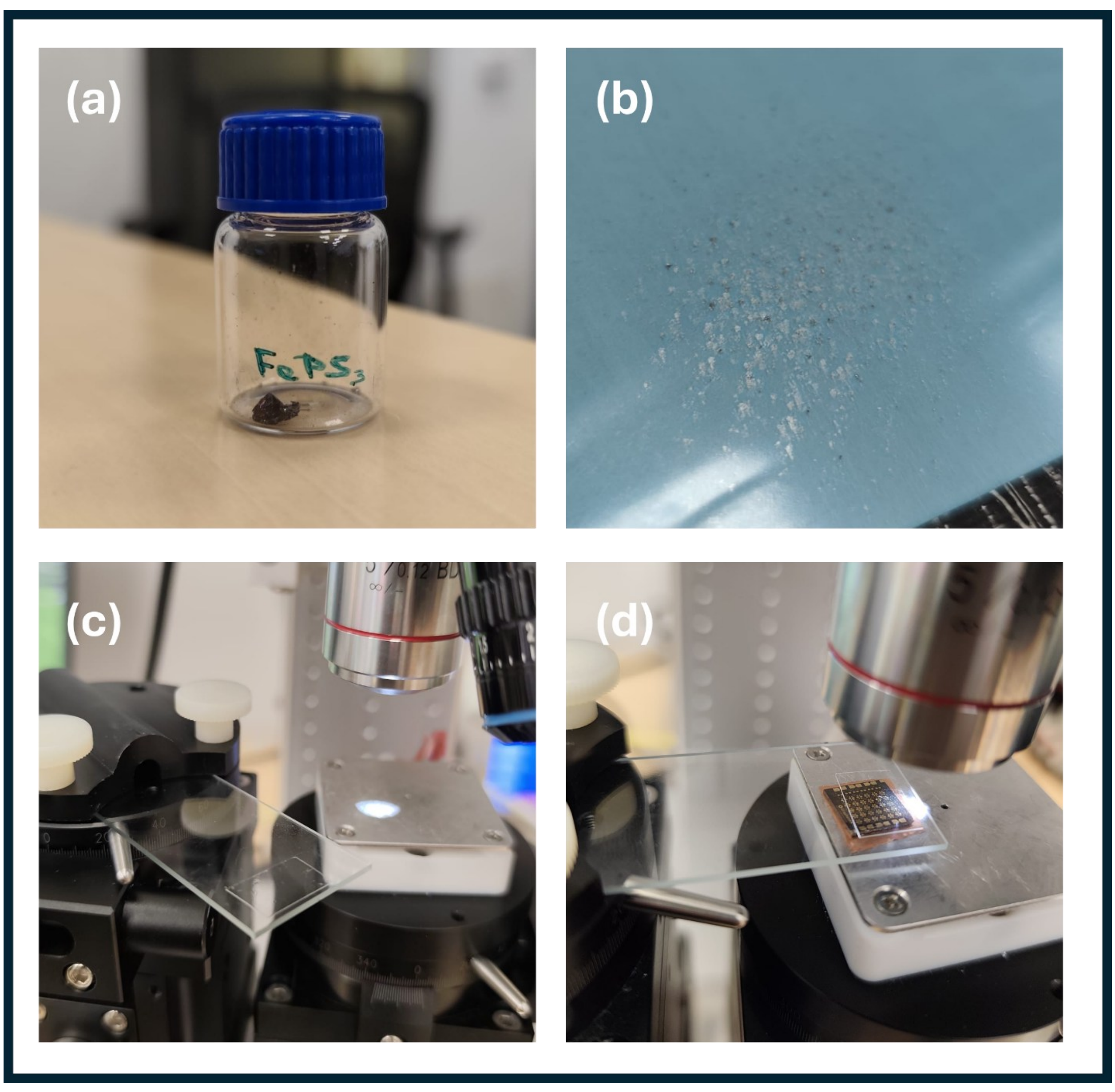

**Figure S1**. Summary of the dry transfer process used to fabricate FEPS/GFET heterostructures. (a) A small quantity of bulk material is taken with a needle and (b) exfoliated using Nitto tape. (c)

Some flakes are picked up with a previously prepared porta with a piece of PDMS polymer. (d) With the help of an optical microscope and a set of micromanipulators, a chosen $FePS_3$ flake is placed from the PDMS in the porta into a pristine GFET device.

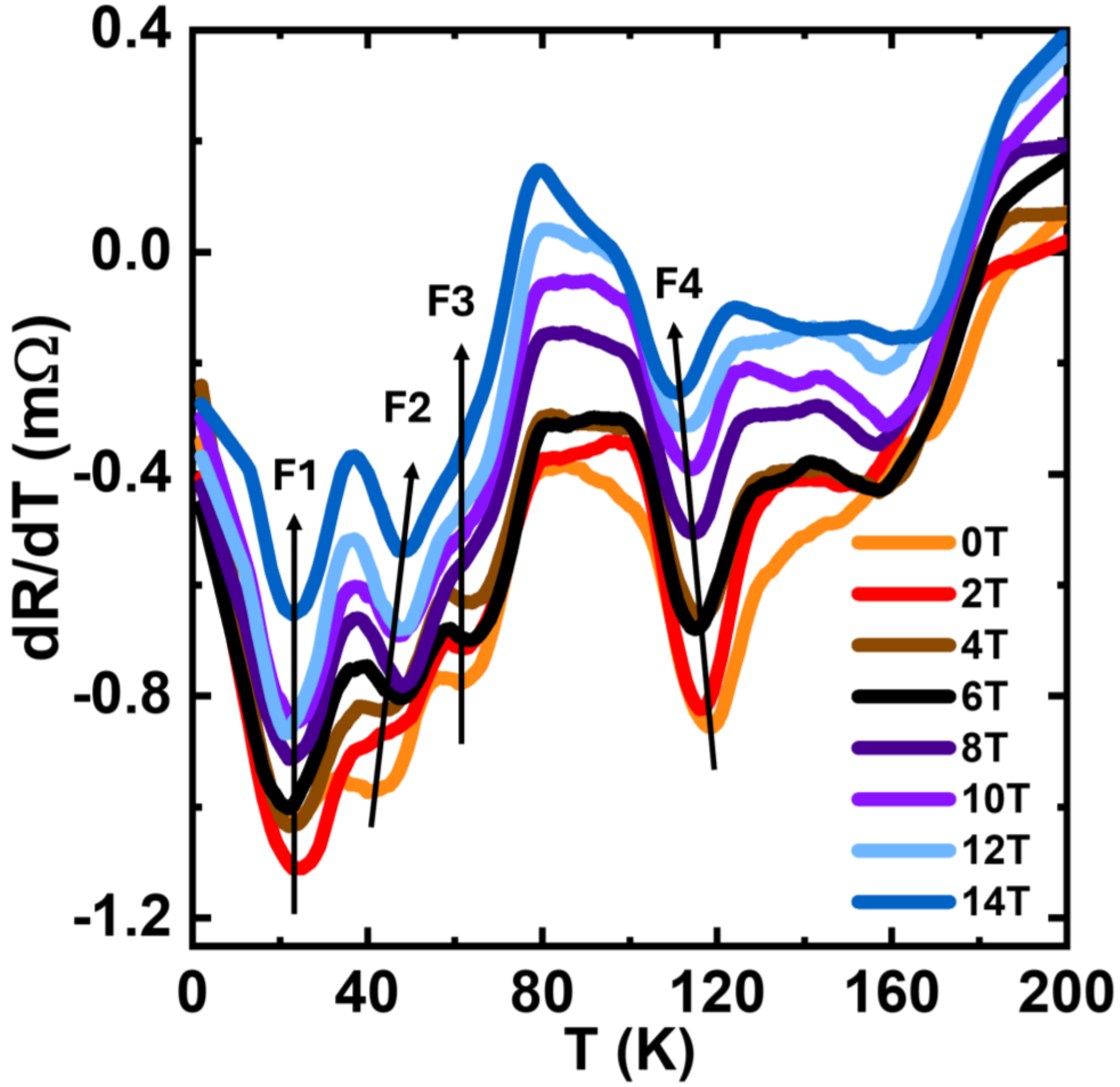

**Figure S2**. Resistance derivative values d$R$/d$T$ as a function of temperature and external magnetic field, calculated from drain-source measurements in the FEPS/GFET heterostructure studied in Figure 2 in the main text (device A). Colour map in the main document has been constructed using these data.

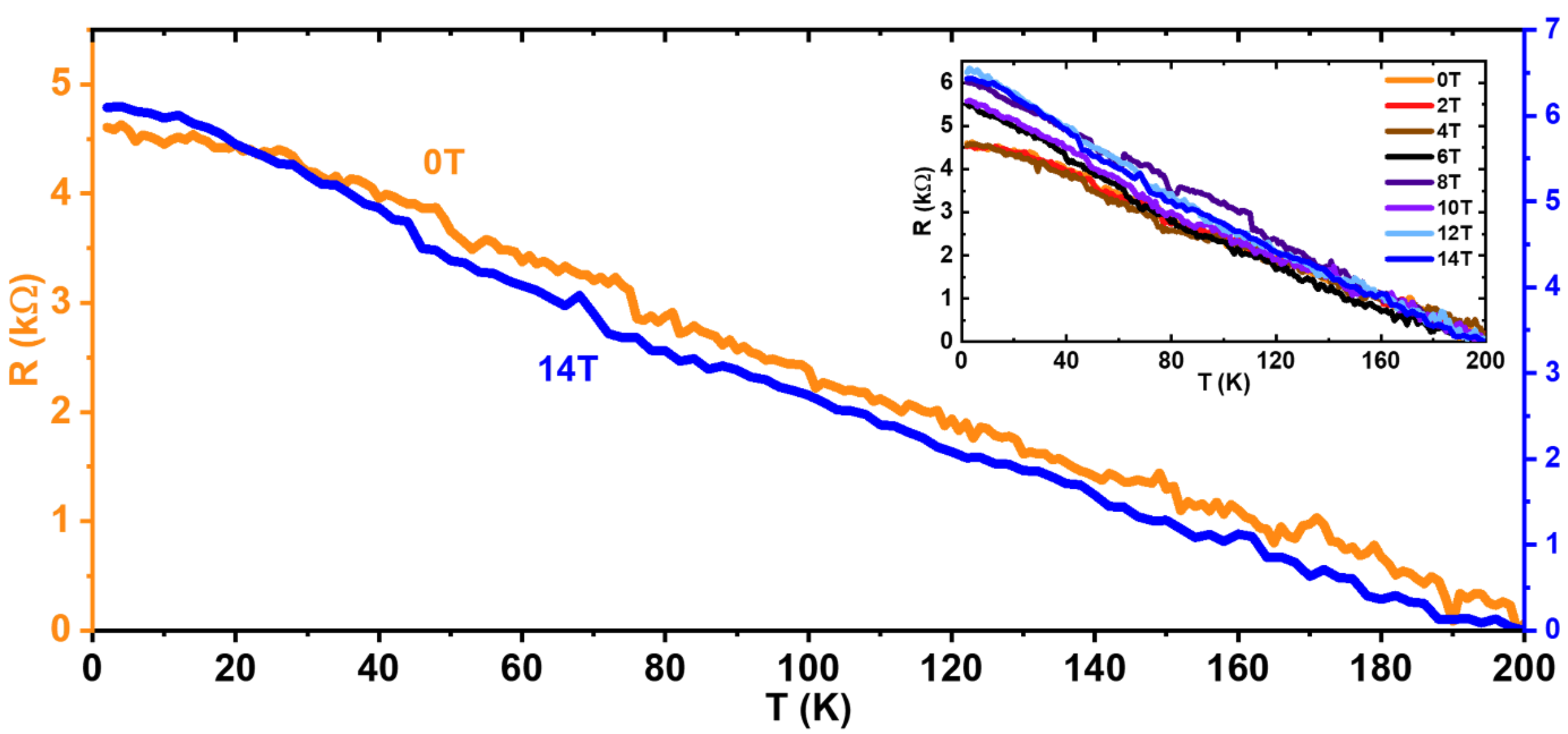

**Figure S3**. (a) Drain-source electric resistance R of a GFET (device B) as a function of temperature T, with applied magnetic field values of 0 T (orange) and 14 T (blue). Inset: additional measurements for 2-12 T.

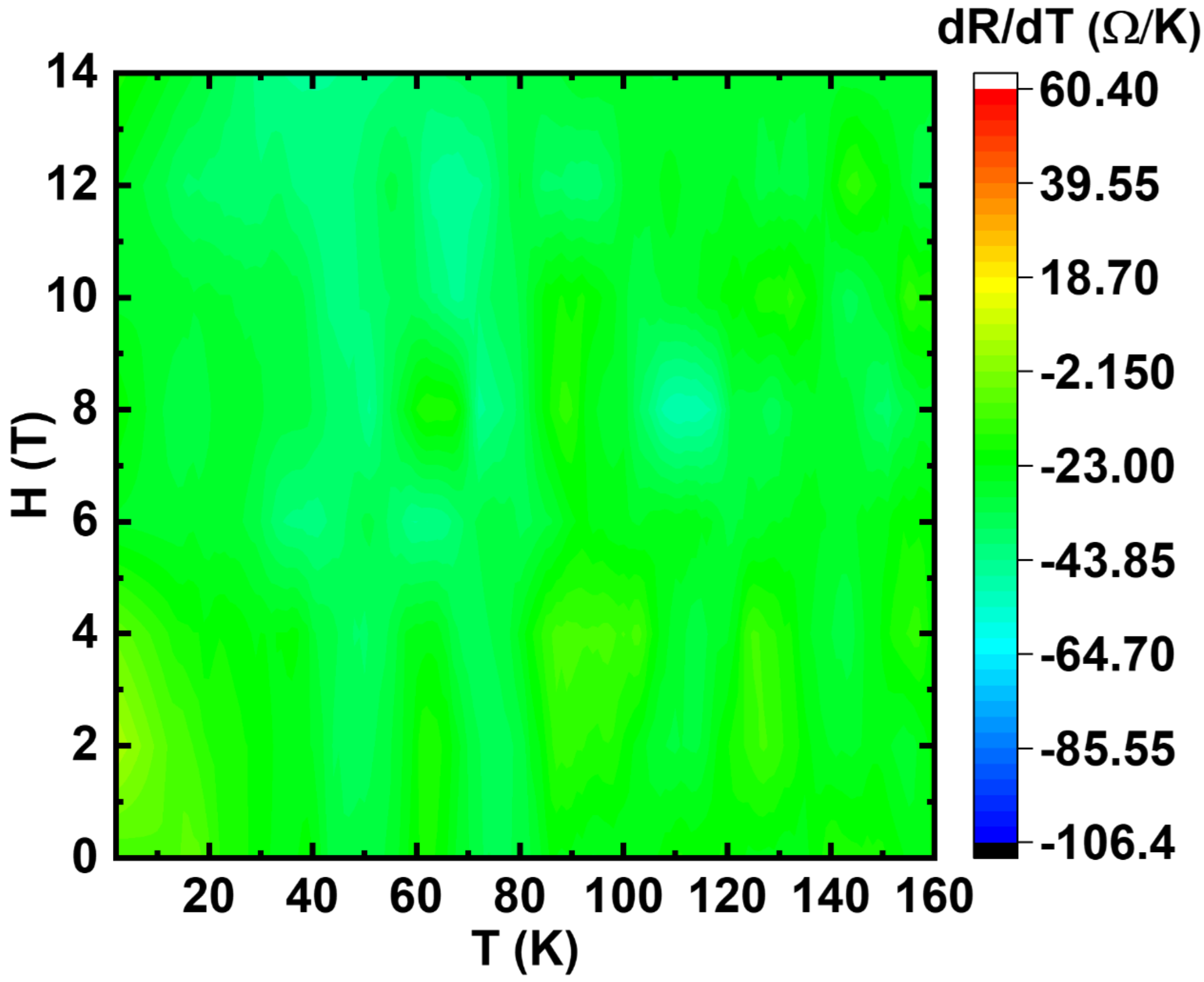

**Figure S4**. Colour map constructed and interpolated from resistance derivative values dR/dT in a GFET device (device B), as a function of both temperature and applied magnetic field.

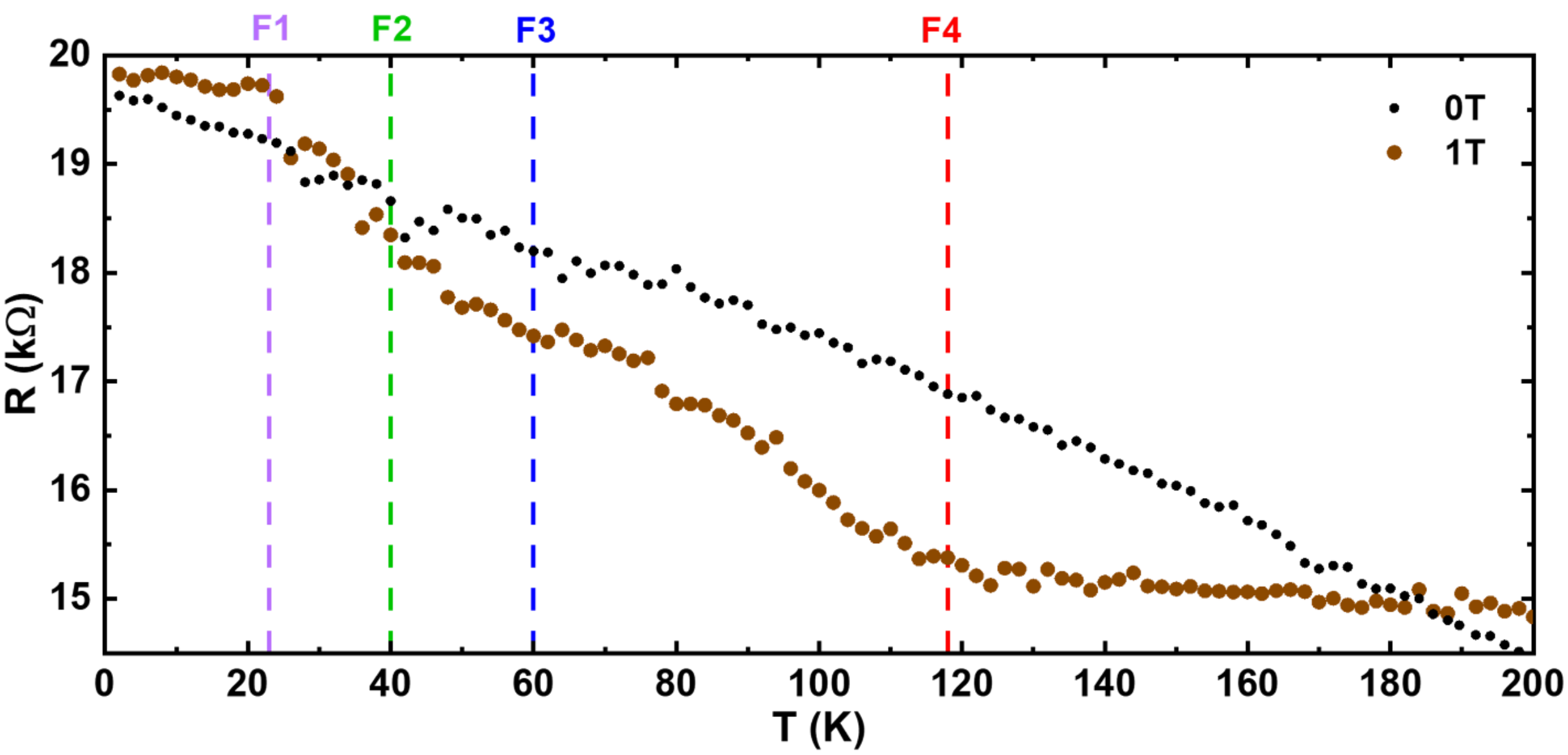

**Figure S5**. Drain-source electric resistance R of an additional FEPS/GFET heterostructure (device C) as a function of temperature T, with applied magnetic field values of 0 T (black) and 1 T (brown). Four blue dashed lines are included as visual help to localize the anomalous features in the curves.

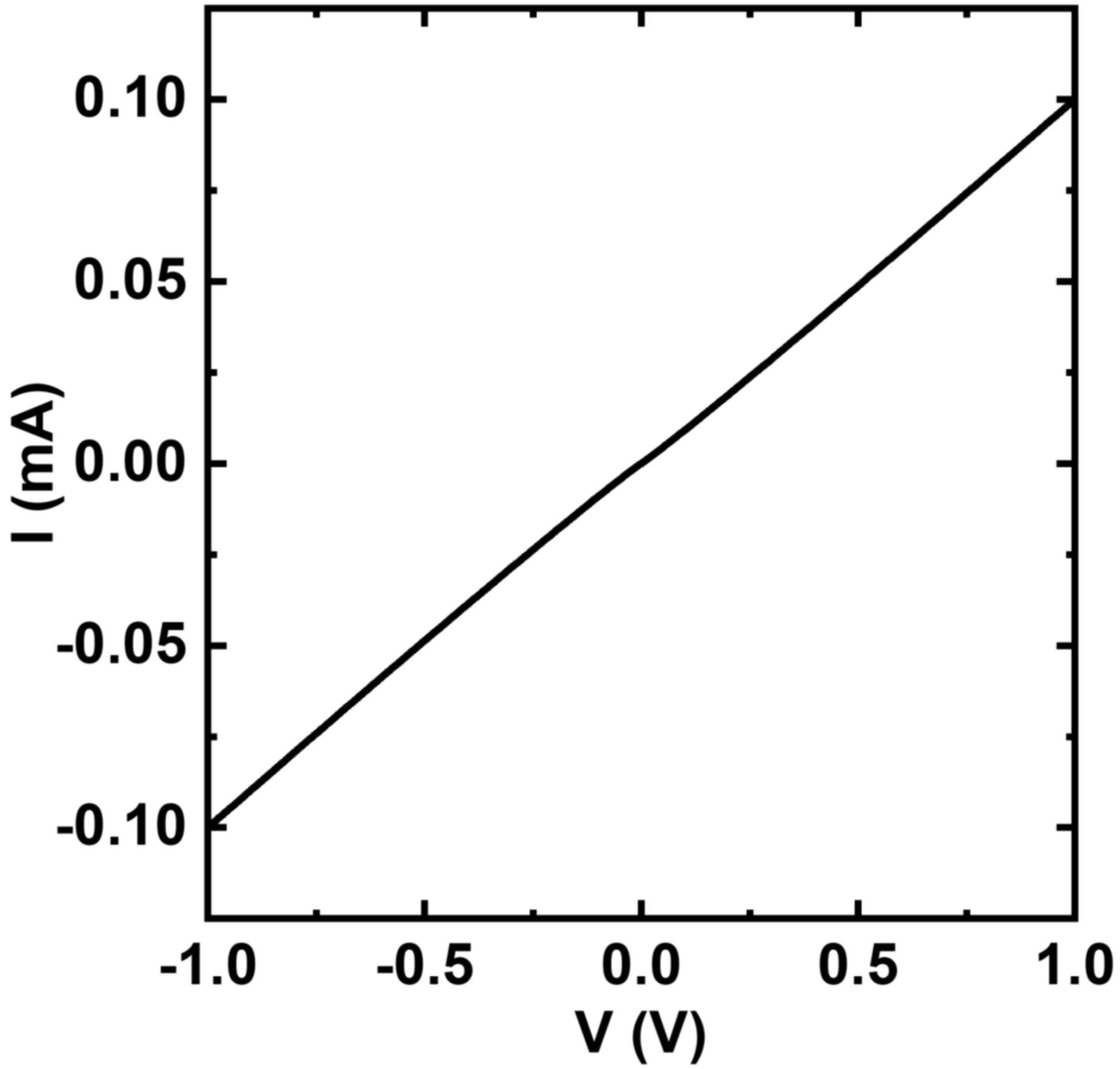

**Figure S6**. IV curve measured for a FEPS/GFET heterostructure (device D) at 3K.

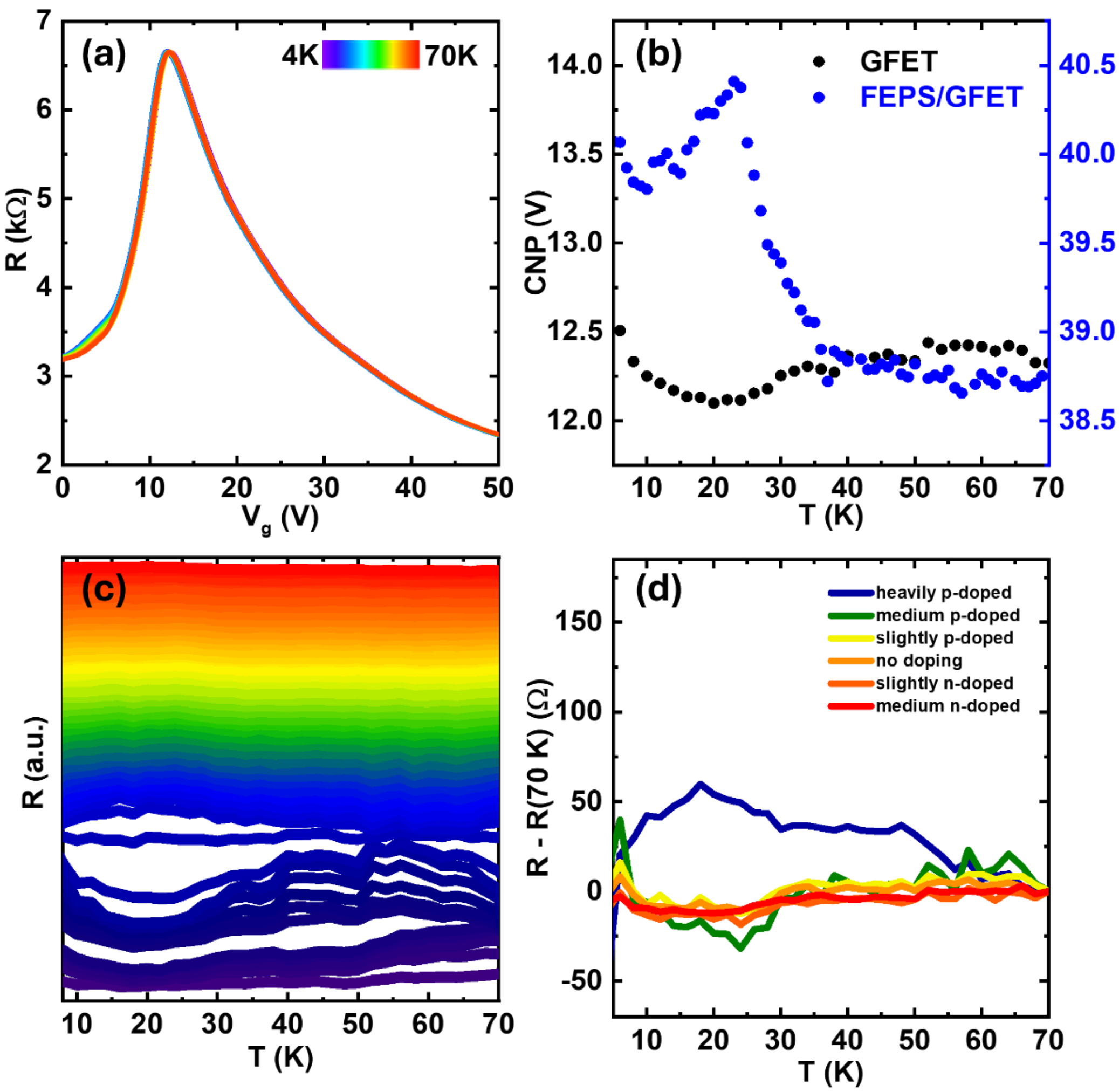

**Figure S7**. (a) Gate voltage traces measured at temperatures from 4 K (blue) to 70 K (red), with a fixed bias voltage of 1 V, for a pristine GFET (device E). (b) Charge neutrality point (CNP) as a function of temperature extracted from (a). CNP curve from FEPS/GFET is included for comparison. (c) Resistance curves as a function of temperature at different gate voltage ranging from 0 V (blue) to 50 V (red). The curves are shifted vertically to facilitate the comparison. (d) Selected curves from (c) corresponding to specific levels of doping. The curves are rescaled to the resistance at the highest temperature for a better comparison.

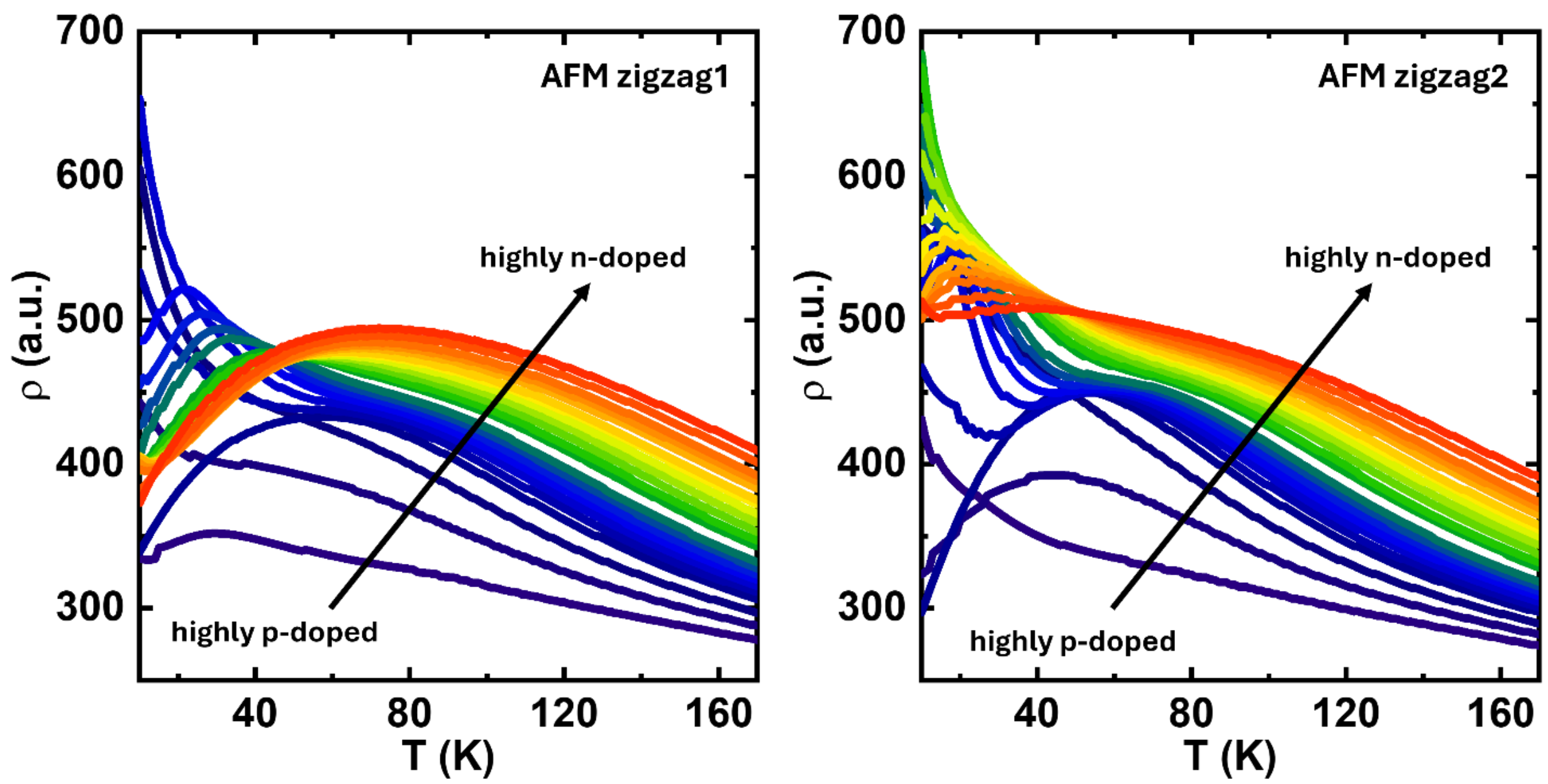

**Figure S8**. Thermal dependence of the resistivity of the FEPS/GFET heterostructure for AFM zigzag1 (a) and AFM zigzag2 (b) phases, with different surface density of n-type and p-type defects. All calculations with optimized geometries resulting from PBE+TS calculations, with U=2.2 eV. Black line corresponds to the resistivity of the heterostructure containing pristine graphene.

**Table S1**. Relative energy of the heterostructures FEPS/GEFT containing the FEPS AFM zigzag1 and zigzag2 phases with different $U_{eff}$ values in PBE+U calculations. All calculations employed the optimized geometries obtained with PBE+U=2.2 eV. $U_{eff}$ and Δ in eV and $\Delta/N_{Fe}$ in eV per Fe.

| $U_{eff}$ | 1.2 | 2.2 | 2.7 | 3.2 |
|---|---|---|---|---|
| $\Delta = E_{zig1/GEFT} - E_{zig2/GEFT}$ | $-1.1 \cdot 10^{-4}$ | $-6.1 \cdot 10^{-4}$ | $-7.3 \cdot 10^{-4}$ | $-4.2 \cdot 10^{-4}$ |
| $\Delta/N_{Fe}$ | $-1.4 \cdot 10^{-5}$ | $-7.6 \cdot 10^{-5}$ | $-9.1 \cdot 10^{-5}$ | $-5.2 \cdot 10^{-5}$ |

**Table S2**. Optimized computational cell parameters for $FePS_3$ layer deposited on graphene for AFM zigzag1 and zigzag2 solutions, compared to those in absence of the substrate from DFT+U=2.2 eV calculations.

| | *a* | *b* | *c* | *α* | *β* | *γ* |
|---|---|---|---|---|---|---|
| FEPS *zigzag1* | 11.9239 | 11.9035 | 20.1605 | 89.97 | 90.04 | 60.06 |
| FEPS *zigzag2* | 11.9007 | 11.9265 | 20.0529 | 89.86 | 89.94 | 60.05 |
| FEPS *zigzag1*/GEFT | 12.2523 | 12.2506 | 20.0657 | 90.04 | 90.01 | 60.00 |
| FEPS *zigzag2*/GEFT | 12.2523 | 12.2506 | 20.0657 | 90.04 | 90.01 | 60.00 |

**Table S3**. Fe-Fe distances for $FePS_3$ layer deposited on graphene for AFM zigzag1 and zigzag2 solutions, compared to those in absence of the substrate (in parenthesis) from DFT+U=2.2 eV calculations.

| Fe-Fe (Å) | AFM *zigzag1* | AFM *zigzag2* |
|---|---|---|
| Fe1-Fe3 | 3.514 (3.424) | 3.520 (3.430) |
| Fe1-Fe5 | 3.509 (3.422) | 3.481 (3.402) |
| Fe5-Fe7 | 3.531 (3.426) | 3.615 (3.515) |
| Fe2-Fe4 | 3.599 (3.480) | 3.631 (3.514) |
| Fe2-Fe6 | 3.505 (3.411) | 3.504 (3.414) |
| Fe6-Fe8 | 3.519 (3.430) | 3.544 (3.436) |
| Fe1-Fe2 | 6.001 (5.868) | 6.047(5.885) |
| Fe3-Fe4 | 6.143 (5.981) | 6.137(5.948) |
| Fe5-Fe6 | 6.016 (5.876) | 5.995(5.861) |
| Fe7-Fe8 | 6.158 (5.989) | 6.131(5.949) |
| Fe1-Fe6 | 3.535 (3.432) | 3.510 (3.424) |

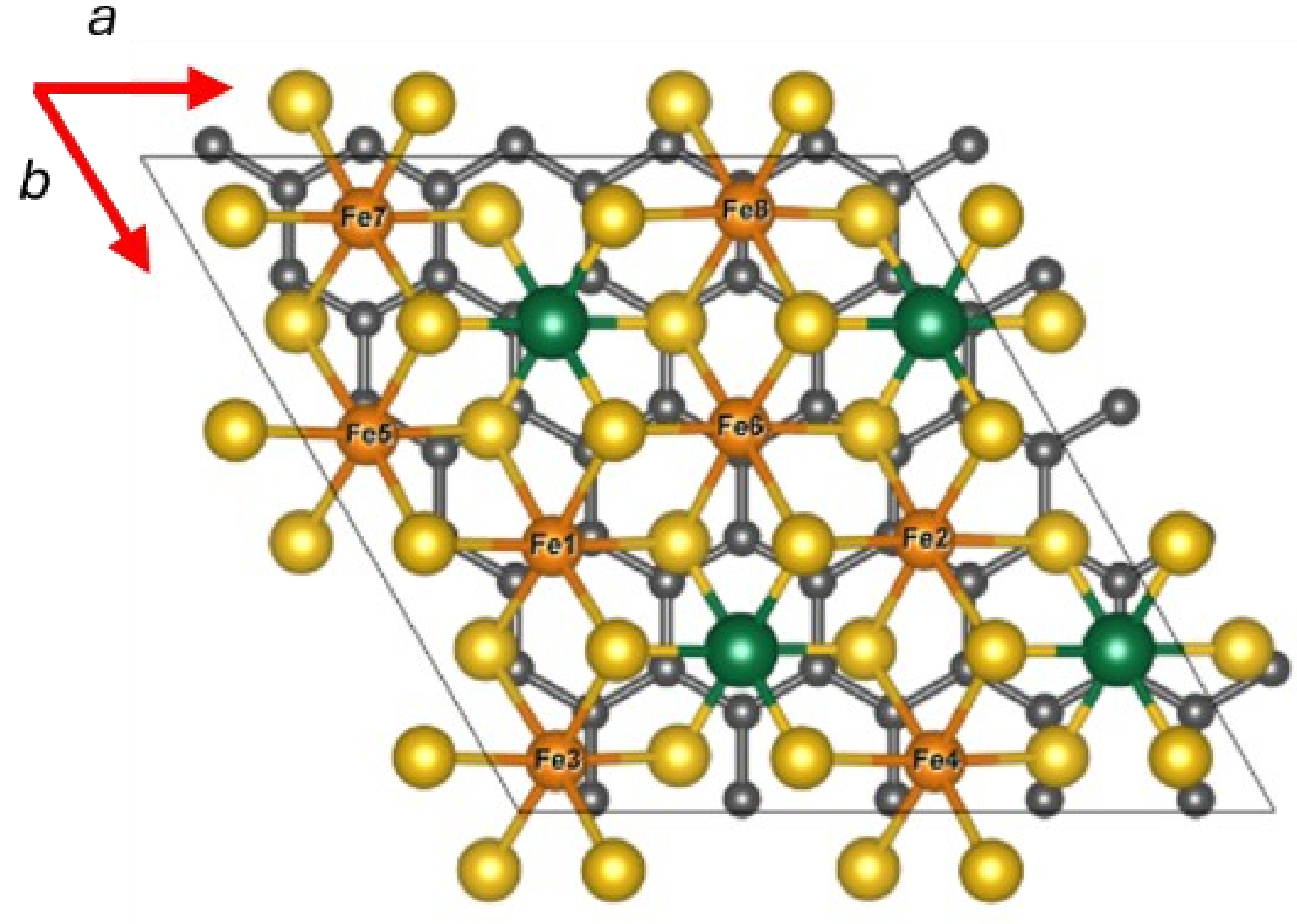

## Experimental section

### Materials

$FePS_3$ crystals were acquired from HQ Graphene (> 99.99% purity, typical lateral sizes of 5 mm).[Error! Reference source not found.] Commercial graphene field-effect transistors (GFETs) with different graphene channel sizes were obtained from Graphenea.[Error! Reference source not found.] For the experiments described in this work, the dimensions of the graphene window vary between 30-50 μm x 30-50 μm.

### Fabrication of heterostructures

Exfoliated FEPS flakes were deposited onto GFETs using a transfer station (HQGraphene) equipped with an optical microscope and a set of micromanipulators. Prior to use, an annealing of the GFET (<300 °C, 2 h) is carried out in high vacuum conditions to ensure the cleanest surface possible.

### Electron transport measurements

Electron transport measurements on pristine graphene and FEPS/GFET heterostructures were performed in a Physical Properties Measurement System (PPMS) 14T from Quantum Design (QD). Temperature was controllably swept from 200 to 3 K using a closed flow of helium. Magnetic fields from 0 to 14 T were applied using a superconductor coil. All measurements were performed in vacuum conditions (~$10^{-6}$ mbar).

Gate-dependent electron transport measurements on pristine graphene and FEPS/GFET heterostructures were performed in a Cryogen-Free Closed Cycle 4K Cryostat from ICE. Temperature was controllably swept from 200 to 4 K by combining the cryostat’s two stage Gifford-McMahon (GM) cooler with a PID-based Lakeshore temperature controller. Gate voltages were applied with a TENMA source. A Keithley 2450 Sourcemeter was employed for both applying drain–source voltages and measuring drain–source current. All measurements were performed in vacuum conditions (~$10^{-6}$ mbar).

## Computational Details

The interaction between graphene and $FEPS_3$ in AFM zigzag1 and zigzag2 phases was studied by means of periodic DFT calculations using VASP code **[3-6],** with the same conditions as in our previous study **[7].**

The $FEPS_3$ flakes are modeled with a single layer, assuming that the layer in contact with the graphene is responsible for the modulation of the graphene transport properties. A 2x2 supercell 11.9 x 11.9 Å, containing 8 Fe atoms is used to model the single layer of $FePS_3$. Graphene is represented with a supercell containing 50 carbon atoms 12.3 x 12.3 Å. The model of the FEPS/GEFT heterostructure was built by means of the *InterfaceBuilder* module of the QuantumWise ATK software package **[8],** maximizing the epitaxial interaction and minimizing the absolute stress between the $FePS_3$ layer and the graphene sheet. The size of the computational cell is 12.25 x 12.25 Å, with a thick vacuum layer of 20 Å in z direction to avoid the interaction between the layers. The complete description can be found in Ref. 7. The heterostructure containing the AFM zigzag2 phase of $FePS_3$ in contact with graphene has been built starting from the optimized structure of the AFM zigzag1, previously reported **[7]**. The so-resulting heterostructure has been optimized with the same criteria than in our previous study (geometry optimizations of both the supercell lattice parameters and the atomic positions employing an 8x8x1 Monkhorst-Pack k-point mesh, with a convergency criterion of 10-6 eV in the electronic relaxation and 0.015 eV $Å^{-1}$ on the Hellmann–Feynman forces). A final single-point calculation with a 21x21x1 Monkhorst-Pack k-point mesh on the optimized geometry is employed to extract the density of states and the relevant information for the transport calculations. The representation of the optimized structures and electronic density is done with VESTA code **[9]**.

The generalized gradient approximation (GGA) is employed with the Perdew–Burke–Ernzerhof (PBE) exchange-correlation functional **[10]** and projector-augmented wave (PAW) potentials **[11,12]**. The valence electrons are described using a plane-wave basis set with a cut-off of 500 eV. An effective Hubbard term ($U_{eff}$) has been used to describe the localized 3d orbitals of Fe centers, using Dudarev approach **[13]**. Van der Waals interactions are considered through the Tkatchenko–Scheffler method **[14]**. Results reported in the main text refer to the $U_{eff}$=2.2 eV value, as in our previous study. Additionally, $U_{eff}$ values from 1.2 to 3.2 eV were tested, in line with those proposed by other groups (Petska et al **[15]** and Amirabbasi and Kratzer **[16]**). As shown in Table S1, the relative stability of the heterostructures containing the AFM *zigzag1* and *zigzag2* phases of $FePS_3$ is almost independent of the adopted $U_{eff}$ value.

The temperature dependent conductivity $\sigma$ of the FEPS/GFET heterostructure in terms of relaxation time τ, was computed employing the BoltzTrap2 code **[17]**, based on the semiclassical Boltzmann transport theory, using the information provided by the previous periodic calculations for each magnetic solution, namely the density of states and eigenvalues. In this method, the transport coefficients are calculated on the frame of the rigid band approximation, which assumes that changing the temperature ($T$), or the Fermi energy of the system ($\mu$), does not change the band structure. The electrical conductivity is computed by means of a linearized version of the Boltzmann transport equation, under constant relaxation time approximation **[16]**:

$$\frac{\sigma(\mu,T)}{\tau} = q^2 \int \sigma^*(\varepsilon,T)\left(-\frac{\partial f^{(0)}(\varepsilon,\mu,T)}{\partial \varepsilon}\right) d\varepsilon$$

where $q$ is the electron charge, $f^{(0)}$ is the Fermi distribution function and $\sigma^{*}(\varepsilon, T)$ is the transport distribution function. The resistivity is finally approximated by the inverse of the trace of the conductivity tensor matrix, assuming a relaxation time of $\tau = 1$ ps for graphene (usual values range from 10 fs to 1 ps).

The transport properties have been evaluated for pristine graphene and considering specific doping levels. For a unit cell with volume $V_c$, the doping is defined as $D_n - D_{int}$, the difference between the intrinsic electronic density, $D_{int} = N_{int}/V_c$, and the density once a certain fraction of electrons, $\delta e$, has been removed (p-doping) or added (n-doping), $D_n = (N_{int} \pm \delta e)/V_c$. The so-resulting $D_n - D_{int}$ values are multiplied by the unit cell $c$ parameter ($2 \cdot 10^{-7}$ cm) to obtain the surface density of defects (in $cm^{-2}$), p- or n-defects.

**References**

1 *HQ Graphene*, $FePS_3$, https://www.hqgraphene.com/FePS3.php

2 *Graphenea*, GFET-S10 for sensing applications, https://www.graphenea.com/products/gfet-s10-for-sensing-applications-10-mm-x-10-mm

3 Kresse, G.; Furthmuller, J. Efficiency of ab-initio total energy calculations for metals and semiconductors using a plane-wave basis set. *Comput. Mater. Sci.* **1996**, *6* (1), 15-50. DOI: https://doi.org/10.1016/0927-0256(96)00008-0

4 Kresse, G.; Furthmuller, J. Efficient iterative schemes for ab initio total-energy calculations using a plane-wave basis set. *Phys. Rev. B* **1996**, *54* (16), 11169-11186. DOI: https://doi.org/10.1103/PhysRevB.54.11169

5 Kresse, G.; Hafner, J. Ab Initio Molecular-Dynamics Simulation of the Liquid-Metal-Amorphous Semiconductor Transition in Germanium. *Phys. Rev. B* **1994**, *49* (20), 14251-14269. DOI: https://doi.org/10.1103/PhysRevB.49.14251

6 Kresse, G.; Hafner, J. Ab initio molecular dynamics for liquid metals. *Phys. Rev. B* **1993**, *47* (1), 558-561. DOI: https://doi.org/10.1103/PhysRevB.47.558

7 García-Martín, A., Sánchez-De-Armas, R., Gullace, S., Calle, E., Martín-Pérez, L., Montenegro-Pohlhammer, N., Jaafar, M., Sanchez Costa, J., Calzado, C. J., & Burzurí, E. Conductance modulation in graphene/antiferromagnet van der Waals heterostructures induced by magnetic order. Nanoscale, **2025**, *17* (35), 20327–20337. DOI: https://doi.org/10.1039/d5nr02538a

8 Smidstrup,S.; Stokbro,K.; Blom, A.; Markussen,T.; Wellendorff, J.; Schneider, J.; Gunst,T.; Verstichel,B.; A Khomyakov, P.; Vej-Hansen,U.G.; Brandbyge,M. et al. QuantumATK: An Integrated Platform of Electronic and Atomic-Scale Modelling Tools, *J. Phys: Condens. Matter* **2020**, *32* (1), 015901. DOI: https://doi.org/10.1088/1361-648X/ab4007

9 Momma, K.; Izumi, F. VESTA 3 for three-dimensional visualization of crystal, volumetric and morphology data. *J. Appl. Cryst.* **2011**, *44* (6), 1272-1276. DOI: https://doi.org/10.1107/S0021889811038970

10 Perdew, J. P.; Burke, K.; Ernzerhof, M. Generalized gradient approximation made simple. *Phys. Rev. Lett.* **1996**, *77* (18), 3865-3868. DOI: https://doi.org/10.1103/PhysRevLett.77.3865

11 Blochl, P. E. Projector augmented-wave method. *Phys. Rev. B* **1994**, *50* (24), 17953-17979. DOI: https://doi.org/10.1103/PhysRevB.50.17953

12 Kresse, G.; Joubert, D. From ultrasoft pseudopotentials to the projector augmented-wave method. *Phys. Rev. B* **1999**, *59* (3), 1758-1775. DOI: https://doi.org/10.1103/PhysRevB.59.1758

13 Dudarev, S. L.; Botton, G. A.; Savrasov, S. Y.; Humphreys, C. J.; Sutton, A. P. Electron-energy-loss spectra and the structural stability of nickel oxide: An LSDA+U study. *Phys. Rev. B* **1998**, *57* (3), 1505-1509. DOI: https://doi.org/10.1103/PhysRevB.57.1505

14 Tkatchenko, A.; Scheffler, M. Accurate Molecular Van Der Waals Interactions from Ground-State Electron Density and Free-Atom Reference Data. *Phys. Rev. Lett.* **2009**, *102* (7)*, 073005*. DOI: https://doi.org/10.1103/PhysRevLett.102.073005

15 Pestka, B.; Strasdas,J.; Bihlmayer, G.; Krzysztof Budniak,A.; Liebmann,M.; Leuth,N.; Boban,H.; Feyer,V.;Cojocariu,I.; Baranowski,D.; Mearini,S.; Amouyal,Y.; Waldecker, L.; Beschoten,B.; Stampfer, C.; Plucinski, L.; Lifshitz, E.; Kratzer,P.; Morgenstern,M. Identifying Band Structure Changes of FePS3 across the Antiferromagnetic Phase Transition, *ACS Nano* **2024**, *18* (47), 32924−32931. DOI: https://doi.org/10.1021/acsnano.4c12520

16 Amirabbasi, M.; Kratzer, P. Orbital and magnetic ordering in single-layer FePS3: A DFT+U study. *Phys. Rev. B* **2023**, *107* (2), 024401. DOI: https://doi.org/10.1103/PhysRevB.107.024401

17 Madsen, G.K.H.; Carrete, J.; Verstraete, M.J. BoltzTraP2, a program for interpolating band structures and calculating semi-classical transport coefficients, *Comput. Phys. Commun*., **2018**, *231*, 140–145. DOI: https://doi.org/10.1016/j.cpc.2018.05.010

18 Singh, D.; Mazin, I.I. Calculated thermoelectric properties of La-filled skutterudites. *Phys. Rev. B*. **1997**, *56* (4), R1650-1653. DOI: https://doi.org/10.1103/PhysRevB.56.R1650